\documentclass[a4paper,11pt,onecolumn]{article}
\pdfoutput=1
\usepackage{jheppub}
\usepackage[T1]{fontenc}
\usepackage{mlmodern}
\usepackage{graphicx}% Include figure files
\usepackage{dcolumn}% Align table columns on decimal point
\usepackage{bm}% bold math
\usepackage{amsmath}
\usepackage{amssymb}
\usepackage{slashed}
\usepackage{accents}
\def\d{d\kern-.8 ex\vrule height 1.3 ex depth-1.24 ex width .7 ex \kern .15 ex}
\def\D{D\kern-1.7 ex\vrule height .87 ex depth-.8 ex width .7 ex \kern .95 ex}

\newcommand{\be}{\begin{equation}}
\newcommand{\ee}{\end{equation}}
\newcommand{\bea}{\begin{eqnarray}}
\newcommand{\eea}{\end{eqnarray}}

\title{Trapping, chaos and averaging in bubbling AdS spaces}

\author[a,b,c]{David Berenstein,}
\author[d]{Mihailo \v{C}ubrovi\'c}
\author[d]{and Vladan {\D}uki\'c}
\affiliation[a]{Department of Physics, University of California, Santa Barbara, CA 93106, USA}
\affiliation[b]{
 Institute of Physics, University of Amsterdam, Science Park 904, PO Box 94485, 1090 GL Amsterdam,
The Netherlands}
\affiliation[c]{ Delta Institute for Theoretical Physics, Science Park 904, PO Box 94485, 1090 GL Amsterdam, The
Netherlands}
\affiliation[d]{Center for the Study of Complex Systems, Institute of Physics Belgrade, University of Belgrade, Pregrevica 118, 11080 Belgrade, Serbia}

\emailAdd{dberenstein@ucsb.edu}
\emailAdd{cubrovic@ipb.ac.rs}
\emailAdd{djukic@ipb.ac.rs}

\date{\today}

\abstract{We discuss chaos and ensemble averaging in 1/2 BPS bubbling $AdS$ spaces of Lin, Lunin and Maldacena (LLM) by studying trapped and escaping null geodesics and estimating their decay rates. We find typical chaotic scattering behavior and confirm the Pesin relation between escape rates, Lyapunov exponents and Kolmogorov-Sinai entropy. On the other hand, for geodesics in coarse-grained (grayscale) LLM geometries (which exhibit a naked singularity) chaos is strongly suppressed, which is consistent with orbits and escape rates averaged over microscopic backgrounds. Also the singularities in these grayscale geometries produce an  attractive potential and have some similarities to  black hole throats trapping geodesics for a long time.
Overall, averaging over the ensembles of LLM geometries brings us closer toward the typical behavior of geodesics in black hole backgrounds, but some important differences remain, in particular the existence of a threshold timescale when the averaging fails.}

\begin{document} 
\maketitle
\flushbottom

\section{Introduction}

The bubbling $AdS$ space solutions of Lin, Lunin and Maldacena (LLM) \cite{Lin:2004nb} are an excellent playground for studying the microscopic physics of smooth geometries, without horizons or singularities, obtained as top-down string theory solutions, and yet with some resemblance to black holes. They  preserve 16 SUSY generators (they are half BPS). The spacetimes are a warped product of a four-dimensional manifold and two three-spheres, and they have an $SO(4) \times SO(4) \times \mathbb R$ spacetime symmetry. These geometries are solutions of type IIB SUGRA, holographically dual to a 1/2 BPS state of $\mathcal N=4$ SYM on $\mathbb{S}^3\times\mathbb{R}$.
The LLM geometries can be completely reconstructed from solving a boundary value problem for a PDE, where the boundary is a two dimensional plane (known as the LLM plane) and the boundary conditions are determined from regularity of the ten dimensional metric. The regularity condition produces a two coloring of the LLM plane.
They can be described as a state in the lowest Landau level (LLL) in a quantum Hall fluid and the dual field theory can be formulated as a quantum mechanical matrix model \cite{Berenstein:2004kk}. In M-theory it is U-dual to the solution of Bena and Warner \cite{Bena:2004jw}. The representation in terms of the LLL of a Fermi fluid yields a very intuitive picture of the solution: the two-dimensional Fermi surface of the fluid divides a plane into black areas (particles) and white areas (holes); the geometry and topology of the black and white patterns classifies completely the solutions \cite{Mosaffa:2006qk}.

As was noted in \cite{Skenderis:2008qn}, LLM geometries are not the best test case for fuzzball geometries, aimed to describe black holes as approximations to smooth string theory solutions, since the corresponding 1/2 BPS ``black hole'' does not have a horizon, but a naked singularity \cite{Myers:2001aq}. The singular generalization, known as the superstar or grayscale solution \cite{Myers:2001aq,Mandal:2005wv,Balasubramanian:2018yjq}, is obtained by generalizing the boundary conditions in the LLM plane: normally, only one of the two three-spheres has vanishing radius in the plane; in superstar/grayscale solutions, the radius of both spheres shrink to zero, which produces a singularity. Crucially, this can be understood as coarse-graining over small-scale details of black and white patterns of solutions, very much in line with the idea of ensemble averaging in gravity \cite{Saad:2019lba,Blommaert:2020seb,Saad:2021rcu,Blommaert:2021etf,Blommaert:2021gha,Blommaert:2021fob}. The purpose of this paper is to explore explicitly if the averaged, grayscale backgrounds also yield ensemble-averaged observable physics. 

%An additional motivation is the analogy with the chaotic billiards problem \cite{Berenstein:2023vtd}, so an interesting question would be \textit{can we reproduce some results from the chaotic billiards theory with fuzzballs and microstate geometries as possible dual gravity descriptions?} See for example \cite{Berry::LesHouches,Berry:1981mom,Berry:1986lxu}.

Despite the fact that the averaged geometry is not a black hole but just a naked singularity, we still hope to capture some interesting phenomenological insights into the question if black holes can be understood as ensemble-averaged objects, or more specifically how to distinguish microstate geometries from black holes. This is in a similar philosophy to \cite{Balasubramanian:2018yjq}, which explores other aspects of the emergent horizon.

Geodesic motion and equations of motion for particles and fields in  black hole spacetimes like the Schwarzschild solution, which is described just by general relativity, are all well known to be integrable (separable), while this is not the case for a generic member of some class of microstate geometries \cite{Frolov:2017kze,Chervonyi:2013eja,Bena:2017upb}. The smoking gun for integrability is the existence of the Killing-Yano conformal tensor \cite{Bena:2017upb,Frolov:2017kze}. It turns out that hidden conformal symmetries, which are responsible for the separability of the wave equation, also allow one to construct the holographic dual CFT description for generic black holes, both supersymmetric and non-supersymmetric ones (and further to use it correctly calculate the black hole entropy via the Cardy formula \cite{Guica:2008mu,Castro:2010fd,Cvetic:2011hp,Cvetic:2011dn}). %%These results suggest that the entropy of any black hole can be calculated in some two-dimensional CFT.

On the other hand, if the smooth microstate geometries are to look like a black hole, we expect the following features that mimic the physics of geometries with horizons: trapping of generic infalling geodesics and diffusion dynamics near the horizon. It is hard to imagine that these can be obtained if the motion of geodesics is integrable in the ensemble of microstate geometries. 
In that case we expect that the motion is quasiperiodic and that the periods of the action-angle variables are not too different from the light-crossing time of the black hole region. 
Chaos (non-integrability) is obviously the key for understanding how generic microstate geometries can both mimic aspects of black holes, and on the other hand differ from them crucially by not having a singularity.
For example, there is a large number of periodic orbits in time independent Hamiltonian chaotic systems, although they are still a set of zero measure.
Other solutions can stay very close to these periodic orbits for a long time and they might seem trapped if the periodic orbit they are close to is ``deep inside the black hole region''. Also, since in chaotic systems we expect mixing in phase space, these can give properties similar to diffusion.

This motivates our study of chaotic features of null geodesics in LLM backgrounds. The non-integrable character of such geodesics is in contrast with the bulk integrability of most black hole backgrounds, while the ''weak'' chaos that is present in the dual CFT description (``BPS chaos'', studied in \cite{Chen:2024oqv}) is again in discrepancy with the maximal chaos of thermal gauge theories, i.e. the well-known fact that black holes are the fastest scramblers, based on the MSS bound $\lambda \leq 2\pi/\hbar \beta$ \cite{Maldacena:2015waa}. On the other hand, even when the bulk dynamics of some simple probes is integrable, for example in the presence of thermal horizons, the MSS bound seems to hold and can be related via the QNM spectrum to the thermalization time scales in the dual gauge theory \cite{Djukic:2023dgk}. This is the case because the Lyapunov exponent is not the best indicator of chaos, since it can also be positive in the presence of an unstable saddle point in an otherwise integrable system. Thus along the lines of that research direction our final goal would be to construct the holographic dictionary for chaos that would be valid regardless of the presence or absence of thermal horizons.
It has been argued indirectly that essentialy all LLM geometries produce chaotic dynamics \cite{Chervonyi:2013eja}, except for the global $AdS_5\times S^5$ ground state. Chaos has already been established numerically for a subset of null geodesics that reside inside the LLM plane in {\em generic} LLM geometries \cite{Berenstein:2023vtd}. These become integrable when one imposes one additional restriction: rotational symmetry in the LLM plane. Hence, it is also interesting to investigate geodesics outside the LLM plane for these setups, which can also be microstates in the grayscale setup. Moreover, one usually imagines that the horizon of the 1/2 BPS black hole when it stops being supersymmetric would form close to the grayscale singular region
of the LLM geometry. The motion of infalling geodesics (especially light) into this black hole requires motion in the direction perpendicular to the LLM plane, so it is important to also study geodesic motion in this extra direction. This paper studies such geodesic motion. We find chaos explicitly, even in situation where the in-plane LLM geodesics become integrable.

Clearly, the phenomenology of chaos is intimately related also to the puzzle of averaging, which is especially appealing as it offers a glimpse into the workings of averaging in higher dimensions. In two- and more recently three-dimensional $AdS$ it was argued that the holographic dual of pure quantum gravity is not one single CFT, but rather an average over an ensemble of CFTs \cite{Saad:2019lba,Saad:2021rcu}. But this feature might well be an artifact of our ignorance about the UV physics \cite{Blommaert:2020seb,Blommaert:2021gha}. Since supergravity is an effective description of string theory, such solutions are not UV-complete and should therefore also display this averaging feature. The grayscale vs. black and white LLM geometries present a perfect testing ground for this reasoning.

The plan of the paper is the following. In Section \ref{secbw} we study the dynamics of geodesics in black and white (''normal'') LLM geometries, and inspect chaos in terms of escape rates, fractal structures and the Pesin relation. In Section \ref{secgray} we perform the same study for the grayscale, singular LLM backgrounds, compare it to the black and white case and show how it can be understood in terms of the ensemble averaging picture. Section \ref{secconc} summarizes the conclusions.

\section{Null geodesics in LLM geometries}\label{secbw}

\subsection{Generalities}

In this section we study the dynamics of null geodesic motion in the LLM geometry. Let us first concisely summarize the basics of the LLM solution \cite{Lin:2004nb,Bena:2004jw}. Its most appealing feature is that, thanks to the dual matrix model \cite{Berenstein:2004kk}, it can be understood as being sourced by a Fermi sea in a two-dimensional plane (the LLM plane), where the inside of the Fermi surface is conventionally considered ``black'', whereas the holes live in ``white'' regions. One can learn a lot from a complete classification of the LLM solutions in terms of black and white patterns. Here we will focus on two representative configurations: disk+ring and 3-disk geometries. We are mainly motivated by \cite{Berenstein:2023vtd} and we extend their analysis from the geodesics in the LLM plane to full 3D dynamics.
This is the residual number of directions that are not  determined by conservations laws.

% $(\xi = 0, P_{\xi} = 0)$ to $\xi \geq 0$ (and $P_{\xi}\neq0$)

The LLM geometry consists of two 3-spheres, $\widetilde S^3$ and $S^3$, and the 3+1-dimensional spacetime with the $x$-$y$ plane (LLM plane) located at $\xi=0$, so that $\xi$ can be interpreted as a radial direction \cite{Lin:2004nb}:
\begin{equation}
    g_{\mu \nu}dx^{\mu} dx^{\nu} = \frac{1}{h^2} \left[ -\left( dt + V_a dx^a \right)^2 + h^4 \left( d\xi^2 + dx_a dx^a \right) + \left( \frac{1}{2} - z \right)d\widetilde\Omega_3^2  + \left( \frac{1}{2} + z \right)d\Omega_3^2 \right],\label{eqllmg}
\end{equation}
with 
\begin{equation}
    h^2 = \frac{1}{\xi} \sqrt{\frac{1}{4}-z^2}, \quad \partial_{\xi} V_a = \frac{\epsilon_{ab} \partial_b z}{\xi}.\label{eqllmh}
\end{equation}
 Notice that the radii of the spheres are given by 
 \begin{equation}
     R_1^2= h^{-2} \left( \frac 12 -z\right), \quad R_2^2= h^{-2} \left( \frac 12 +z\right)
 \end{equation}
and that therefore $R_1^2 R_2^2= h^{-4} (1/4 -z^2)= \xi^2$. Thus when $\xi\to 0$ necessarily at least one of the spheres degenerates to zero size.

We thus have a family of solutions determined by the function $z(\xi,x,y)$. The equation is obtained from compatibility of the various Killing spinor equations that must be satisfied to preserve the supersymmetries
\begin{equation}
    \partial_a \partial_a z + \xi \partial_{\xi} \left( \frac{\partial_{\xi}z}{\xi} \right) = 0.\label{eqllmz}
\end{equation}
This function $z$ only depends on the three variables $x,y,\xi$, so the time direction $t$ has translation symmetry. Given the conservation laws, only motion in $x,y,\xi$ will need to be solved explicitly.
For the geometry to be smooth (devoid of singularities) one of the spheres has to vanish in the LLM plane. This leads to the ``black and white`` picture of the boundary conditions at $\xi = 0$ for the Killing spinor equation:
\bea
\lim_{\xi \rightarrow 0} z = +1/2,~~\widetilde S^3\textrm{ contracts to }0\nonumber\\
\lim_{\xi \rightarrow 0} z = -1/2,~~S^3\textrm{ contracts to }0.\label{eqllmbc}
\eea
The regularity condition is that only one sphere shrinks to zero at generic point in the LLM plane, so the function $h^2$ stays finite.

\begin{figure}[ht]
    \centering
    (A)\includegraphics[width=.35\linewidth]{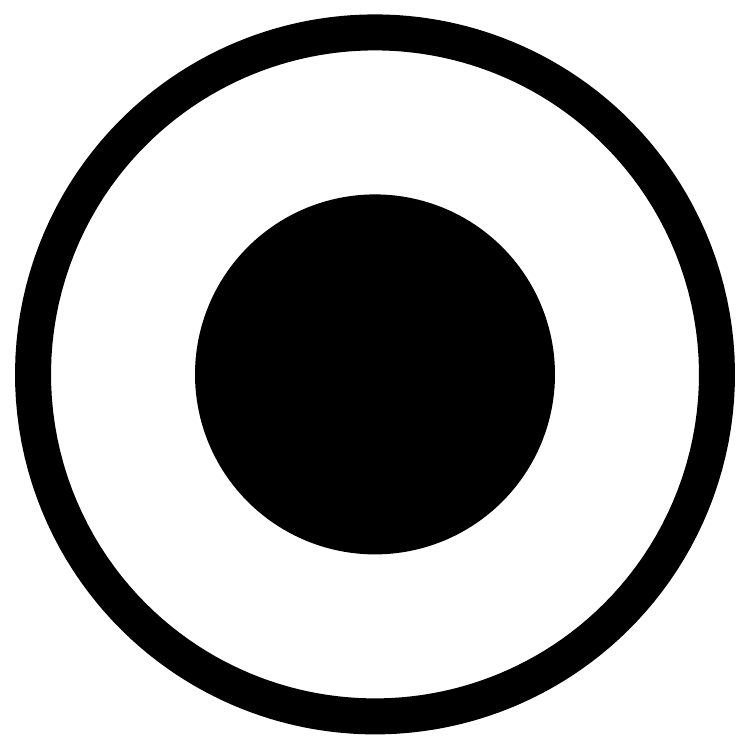}
    (B)\includegraphics[width=.55\linewidth]{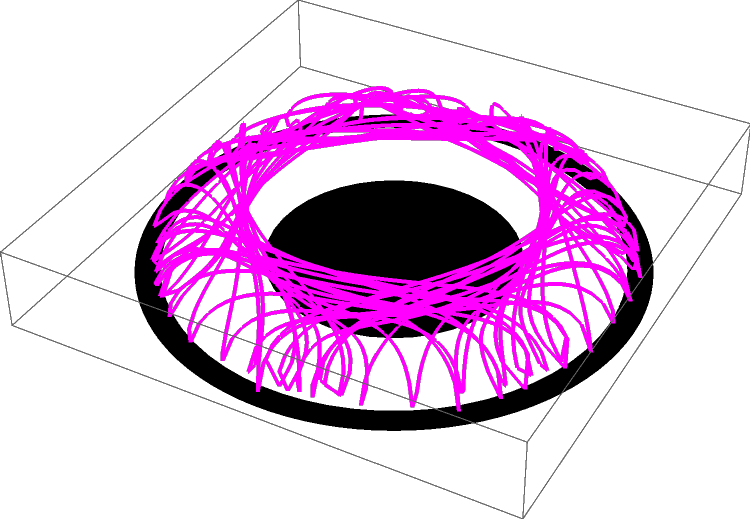}
    \caption{Disk+ring configuration in the LLM plane (A) and the null geodesic (magenta) in the whole 3D space ($x,y,\xi$) (B) in the same background.}
    \label{figbackgnddiskring}
\end{figure}

\begin{figure}[ht]
    \centering
    (A)\includegraphics[width=.35\linewidth]{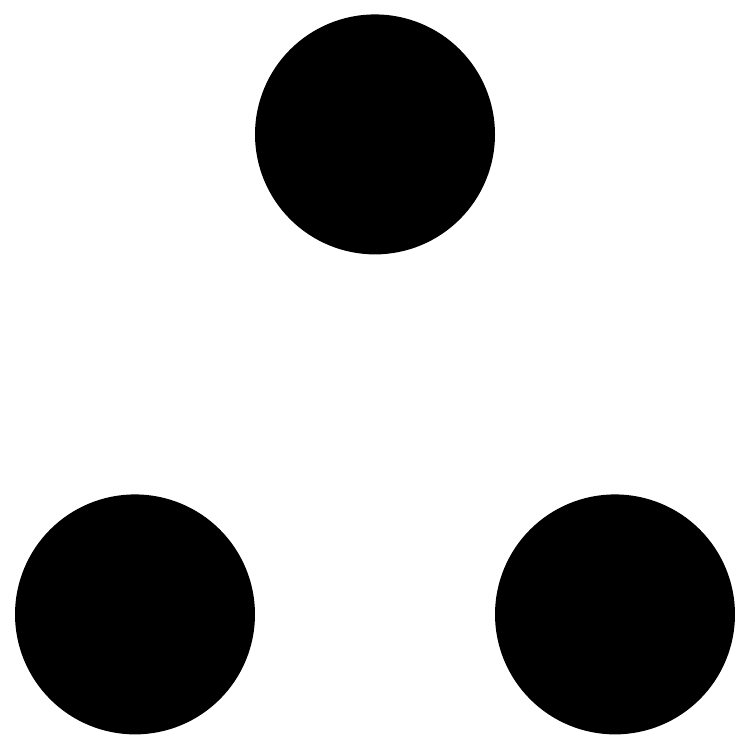}
    (B)\includegraphics[width=.55\linewidth]{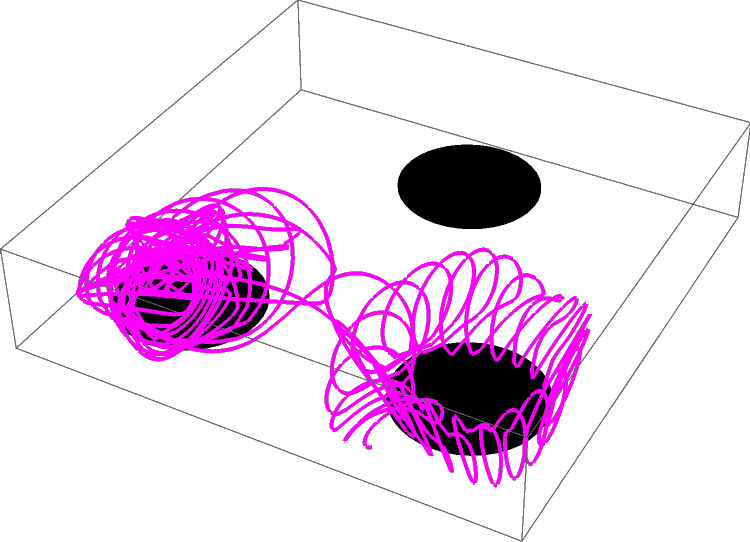}
    \caption{The 3-disk configuration in the LLM plane (A) and a null geodesic (magenta) in the whole 3D space ($x,y,\xi$) (B) in the same background. Notice how the trajectory can jump between disks after being {\em trapped} near one for a while to become trapped by another one.}
    \label{figbackgnd3disks}
\end{figure}

For example, a black disk in the plane would correspond to the $AdS_5 \times S^5$ geometry. We are interested in more complicated solutions where the integrability of probe geodesics breaks down, specifically:
\begin{enumerate}
\item Disk+ring configuration, with a black disk and a black ring concentric with the disk in a white plane (Fig.~\ref{figbackgnddiskring}). This geometry is still integrable in the LLM plane \cite{Berenstein:2023vtd} (because of the conserved angular momentum) but the motion in the full 3+1-dimensional space (not counting the spheres) is chaotic. This configuration is straightforwardly generalized to a system of multiple concentric rings.
\item The 3-disk configuration, a set of 3 black disks in the white plane (Fig.~\ref{figbackgnd3disks}). Unless specified otherwise, we use these same parameters for all 3-disk geometries in the paper (disk radii $R=5$, disks centered at $(x_c,y_c)=(0,0),~(20,0),~(10,20)$).
\end{enumerate}
Both setups can be obtained as superpositions of (black and white) circles in the LLM plane. After all, these circle configurations are a complete classification of the quantum states that preserves half the supersymmetry \cite{Berenstein:2004kk}. 
For disk+rings we prescribe a set of radii $R_1>R_2>\ldots R_n$. For disk+(single) ring, the main case in this section, we have $n=3$ and, unless specified otherwise, the radii are $R_1=5$, $R_2=9$, $R_3=10$. The solution to the Killing spinor equation (\ref{eqllmz}) for a disk of radius $R$ reads
\begin{equation}
    z(\xi,x,y;R) = \frac{\rho^2 + \xi^2 -R^2}{2 \sqrt{(\rho^2 + \xi^2 + R^2 )^2 - 4 \rho^2 R^2}}, \quad \rho^2 \equiv x^2 + y^2.\label{zsinglesol}
\end{equation}
Therefore, the full solution is the superposition
\begin{equation}
    \tilde z_\mathrm{full} = \sum_i (-1)^{i+1} \widetilde z(\xi,x,y;R_i), \quad \tilde z(\xi,x,y;R_i) = z(\xi,x,y;R_i) - 1/2.\label{zfullsol}
\end{equation}
In this case we have cylindrical symmetry, so we can employ the polar coordinates $\{ x = \rho \cos \phi, ~ y = \rho \sin \phi \}$, which brings some additional simplifications: the vector field $V_a$ simplifies to
\begin{equation}
    V_\rho = \cos \phi ~V_x + \sin \phi ~V_y = 0, \quad V_{\phi} = \rho\left(- \sin \phi ~ V_x + \cos \phi ~ V_y\right).
\end{equation}

For the 3-disk solution, besides the set of disk radii $\{ R_1, R_2, R_3 \}$, we also need to specify the positions of disk centers $\{ (x_{01}, y_{01}), (x_{02}, y_{02}), (x_{03}, y_{03}) \}$. The full solution is again the superposition of disks of the form (\ref{zfullsol}), with the same solution for a single disk as in (\ref{zsinglesol}), but now the disk center is not at the origin and all disks are black, therefore:
\begin{eqnarray}
    &&z(\xi,x,y;R,x_0,y_0) = \frac{\rho^2 + \xi^2 -R^2}{2 \sqrt{(\rho^2 + \xi^2 + R^2 )^2 - 4 \rho^2 R^2}}, \quad \rho^2 \equiv (x-x_0)^2 + (y-y_0)^2\nonumber\\
    &&\tilde z_\mathrm{full} = \sum_i \widetilde z(\xi,x,y;R_i,x_{0i},y_{0i}), \quad \tilde z(\xi,x,y;R_i,x_{0i},y_{0i}) = z(\xi,x,y;R_i,x_{0i},y_{0i}) - 1/2~~~~~~~~~~~\label{z3sol}
\end{eqnarray}

Now that we have specified the backgrounds, we can write down the geodesic equations. We start from the Lagrangian for the null geodesic
\begin{equation}
    \mathcal L = \frac{1}{2} ~ g_{\mu \nu}\left[x(\tau)\right] \dot x^{\mu}(\tau) \dot x^{\nu}(\tau).\label{eqLa}
\end{equation}
Due to the $SO(4) \times SO(4) \times \mathbb R$ symmetry, we have a set of three integrals of motion that we label as $\{J_-,J_+, E\}$, following \cite{Berenstein:2023vtd}. The Hamiltonian in the general form reads
\begin{equation} \label{eqHa}
    \mathcal H = \frac{1}{2 h^2}\left[ P_{\xi}^2 + \left( P_x + E V_x \right)^2 + \left( P_y + E V_y \right)^2 - h^4 \left( E^2 - \frac{2J_-^2}{1 - 2z} - \frac{2J_+^2}{1 + 2z} \right) \right],
\end{equation}
where $P_\xi,P_x,P_y$ are the canonical momenta. The quantity $E$
translates to $\Delta -J$ in the CFT, where $\Delta$ is the dimension of the dual operator  to the full quantum state of the system and $J$ is its R-charge. This is the departure from being a 1/2 BPS state, where $J$ includes the background geometry, similar to the BMN Hamiltonian \cite{Berenstein:2002jq}, see also \cite{Berenstein:2020jen}. The corresponding quantum background state is an eigenstate of $\Delta-J$, but not $\Delta,J$ separately.

Since null geodesics are solutions with ${\mathcal H}=0$, and these are invariant under conformal rescalings of the metric, we can also use the modified Hamiltonian given by
\begin{equation} \label{eqHaconf}
    \mathcal H' = \left[ P_{\xi}^2 + \left( P_x + E V_x \right)^2 + \left( P_y + E V_y \right)^2 - h^4 \left( E^2 - \frac{2J_-^2}{1 - 2z} - \frac{2J_+^2}{1 + 2z} \right) \right],
\end{equation}
which makes it easier to understand the trajectories, as the kinetic term is canonical.
The terms $EV_x, EV_y$ produce an effective magnetic field in the directions $x,y,\xi$. Also, away from $z=\pm1/2$ (basically $\xi=0$), the terms in the effective potential with $J_-^2,J_+^2$ are regular on the $(x,y,\xi)$ base. In these geometries $h$ is finite at a generic point on the LLM plane, except at the boundaries between black and white regions. There, the metric is still smooth \cite{Lin:2004nb}, but one needs a change of coordinates to see that it is locally a plane wave geometry. 
Essentially, the term with $J_-^2/(1-2z)$ provide a repulsive force from the region $z=1/2$ and the term with $J_+^2/(1+2z)$ provides a repulsive force from the region $z=-1/2$. 
The trajectories that live inside the LLM plane have either $J_-=0$ or $J_+=0$, so they are very constrained.

\subsection{Disk+ring}

Following \cite{Chervonyi:2013eja}, we find that the LLM geometries which are invariant under rotations in the LLM plane, i.e. $\partial_{\phi}z = 0$, are best dealt with in spherical coordinates:
\begin{equation} \label{eqsphCoords}
    \xi = r \sin \theta, \quad x = r \cos \theta \cos \phi, \quad y = r \sin \theta \sin \phi.
\end{equation}
The Hamiltonian (\ref{eqHa}) in these coordinates reads
\begin{equation} \label{eqHaDRpolar}
  \mathcal H'=  2h^2 ~ \mathcal H =  P_r^2 + \frac{P_{\theta}^2}{r^2} + \frac{(P_{\phi}+E V_{\phi})^2}{r^2 \cos^2 \theta} - h^4 \left( E^2 - \frac{2J_-^2}{1 - 2z} -\frac{2J_+^2}{1 + 2z} \right),
\end{equation}
where $V_\phi = r \cos \theta \left( - \sin \phi ~ V_x + \cos \phi ~ V_y \right)$. The extra integral of motion (the conserved angular momentum) is $P_{\phi}$. We are thus dealing with a system with two degrees of freedom $\{r, \theta \}$, i.e. a four-dimensional phase space.

\begin{figure}[ht]
    \centering
    (A)\includegraphics[width=.45\linewidth]{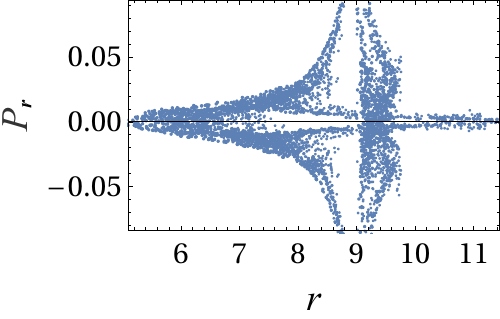}
    (B)\includegraphics[width=.45\linewidth]{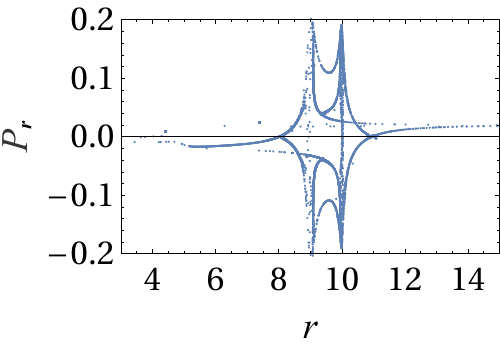}
    \caption{Poincare section $P_{\xi} = 0$ in $P_r-r$ plane for two sets of null geodesic orbits in the LLM geometry sourced by disk + ring `source' pattern in a plane. Integrals of motion are $\{ J_- = 0.01, J_+ = 0.01, E=0.1 \}$ and $P_{\phi} = 0.002$. The panels (A) and (B) differ just by boundary conditions: in (B) we see trapping, where orbits explore a smaller volume in phase space, while remaining equally chaotic.}
    \label{figPS_disc_ring_PrRr}
\end{figure}

In Fig.~\ref{figPS_disc_ring_PrRr} we give the Poincare section defined by $P_{\theta} = 0 ~ (P_{\xi} = 0)$. The areas densely filled with points, clearly visible in the panel (A), correspond to the chaotic sea. In the panel (B) we zoom in into a region with long escape times/low escape rate, corresponding to trapped orbits. Here the chaotic sea is much less prominent although as we will see the Lyapunov exponents are of the same order of magnitude. The orbit in Fig.~\ref{figbackgnddiskring}(B) also gives a qualitative idea of the dynamics.

\begin{figure}[ht]
    \centering
    (A)\includegraphics[width=.45\linewidth]{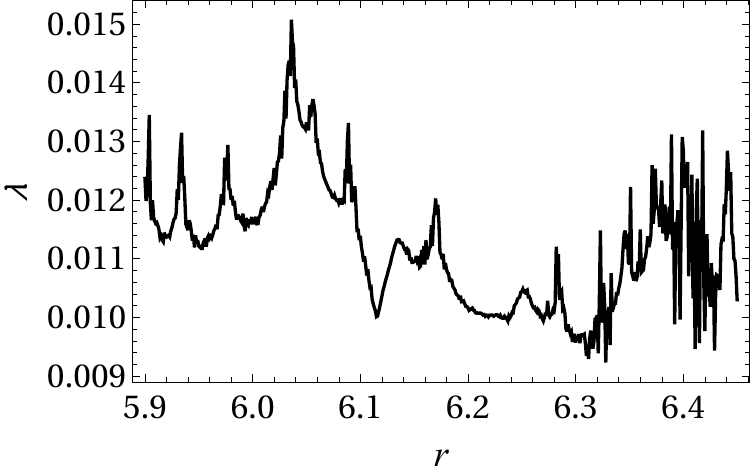}
    (B)\includegraphics[width=.45\linewidth]{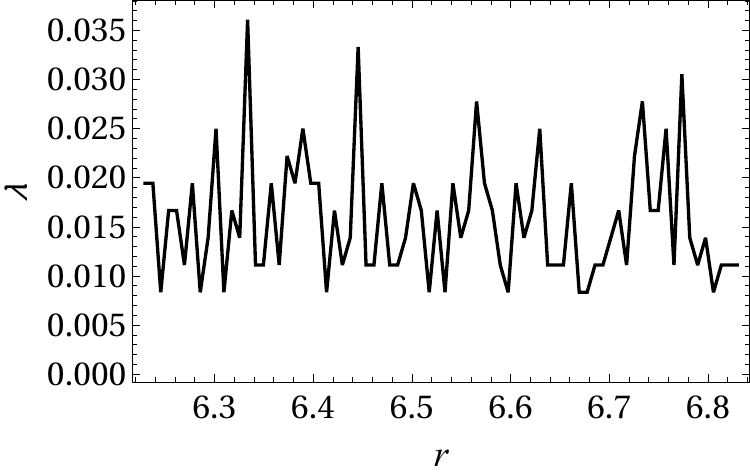}
    \caption{Positive principal Lyapunov exponents for the disk+ring system, with the same initial conditions and integrals of motion as in Fig.~\ref{figPS_disc_ring_PrRr}. The sticky orbits of Fig.~\ref{figPS_disc_ring_PrRr}(B) have about the same Lyapunov exponent (even somewhat larger in this case) as the orbits in Fig.~\ref{figPS_disc_ring_PrRr}(A) which are obviously chaotic.}
    \label{figdiskringLE}
\end{figure}

As a more quantitative chaos indicator and also to check the Pesin theorem, we compute the maximum Lyapunov exponent $\lambda$.\footnote{This is different from the principal Lyapunov exponents $\lambda_i$ obtained as the eigenvalues of the variational matrix.} Similar to \cite{Berenstein:2023vtd}, we find that $\lambda$ is non-zero but extremely small -- Fig.~\ref{figdiskringLE}. This is likely because of the strongly mixed nature of the phase space, with remnants of invariant tori. These are seen as the nearly empty spaces in the Poincare section and will also show up as smooth areas in the escape rate calculation. For some numerical subtleties on computing $\lambda$ see Appendix \ref{secapplambda}.

\begin{figure}[ht]
    \centering
    (A)\includegraphics[width=.45\linewidth]{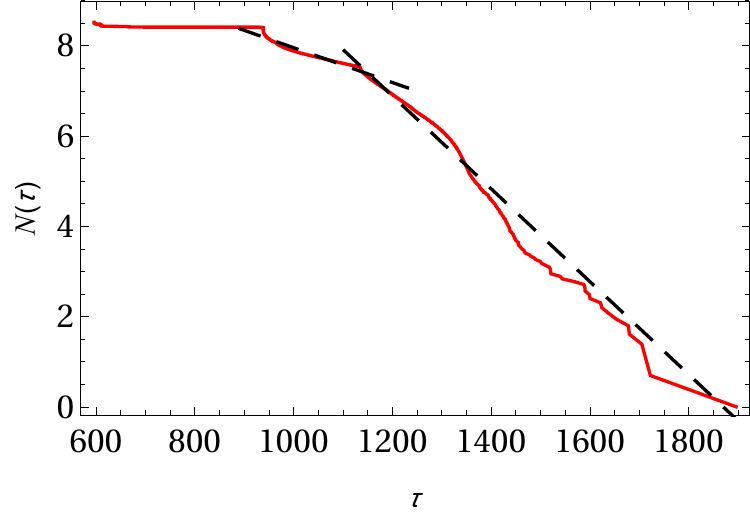}
    (B)\includegraphics[width=.45\linewidth]{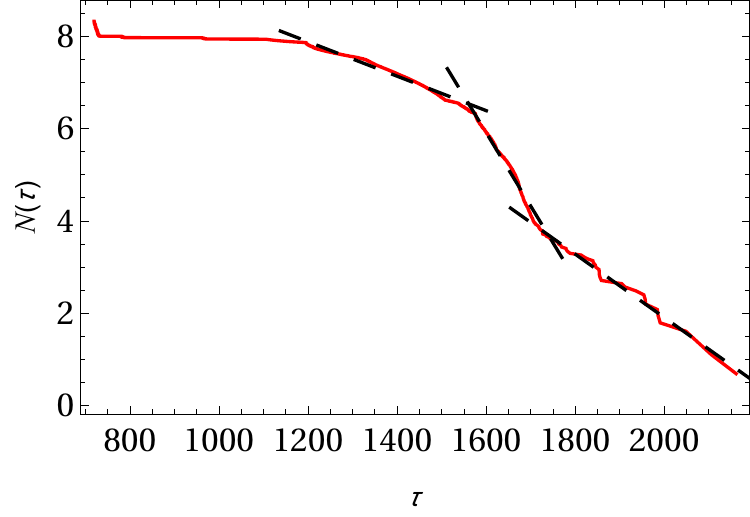}
    (C)\includegraphics[width=.45\linewidth]{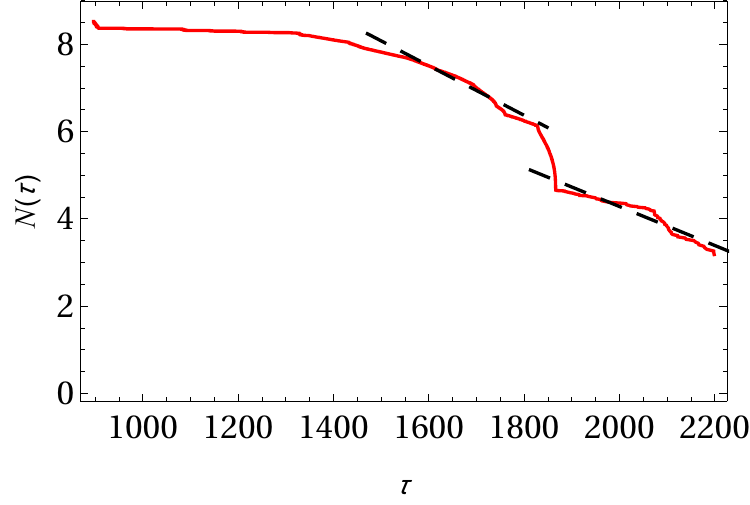}
    (D)\includegraphics[width=.45\linewidth]{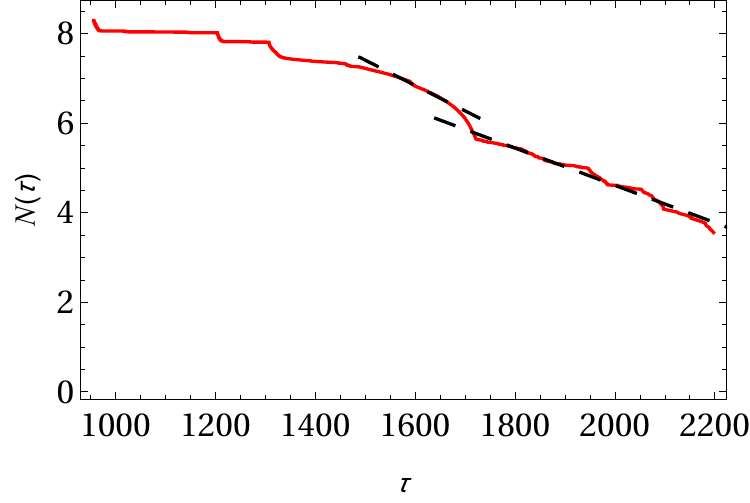}
    \caption{Decay of the number of trapped orbits as a function of proper time $N(\tau)$ (red full lines) for the disk+ring geometry, for four cells of initial conditions of decreasing size, given by $r^{(0)}=20,P_r^{(0)}=-0.02$, $P_\theta^{(0)}=-0.001$, $\phi^{(0)}=\pi/8$ and $\theta^{(0)}$ from the interval $\theta^{(00)}\pm\Delta\theta$, with $\Delta\theta=10^{-3},10^{-7},10^{-8},10^{-9}$ (A to D respectively). The black dashed lines yield the best linear fit, the slope defining the escape rate $\gamma$ from $N(\tau)\sim\exp(-\gamma\tau)$. We detect four distinct populations of orbits, corresponding to the four escape rates on the fits: $\gamma\approx 0.010,0.0038,0.0043,0.0057$. }
    %\caption{Escape rate for a set of null geodesic orbits in the LLM geometry sourced by disk + ring pattern in a plane, in the $r$-direction (A) and in the $\xi$-direction (B). Integrals of motion for this set of orbits are $L = 0.01, \widetilde L = 0.01, E=0.1$ and $P_{\phi} = 0.002$ (A); $L = 0.001, \widetilde L = 0.001, E=0.02$ (B), and the initial conditions are XXX (A) and $\xi^{(0)} = 10, P_{\xi}^{(0)} = -0.01$ (B). The linear fit yields the escape rates $\bar{\gamma}=0.94\times 10^{-4}$ (A) and $\bar{\gamma}=1.84\times 10^{-3}$ (B). In (B), most orbits first evolve towards the LLM plane and get trapped near it for a while before they escape to large $\xi$. \mc{How exactly do we see the trapping? What is the essential difference btw these two plots?}\vd{ plot vs $t(\tau)$ not $\tau$}}
    \label{figGamma_disc_ring1}
\end{figure}

The central quantity for the deeper questions of our study (how closely can LLM geometries and their ensemble averages mimic black holes?) is the escape rate $\gamma$ (see Fig. \ref{figGamma_disc_ring1}).
The rate $\gamma$ is defined for an ensemble of initial conditions and basically measures how many orbits are still inside the trapped region after time $t$.
However, we first need to define an ``escape event''. Strictly speaking, an orbit has escaped to infinity (without ever coming back) once it crosses the Lyapunov curves \cite{Eckmann:1985}. In practice, computing the Lyapunov curves takes time and is not worth the effort: we just define some arbitrary long-distance cutoff in $r$ and check that the result is reasonably robust to changing the cutoff. One is faced with the same problem when studying chaotic matrix models as toy models of black 
holes \cite{Berenstein:2016zgj} and the problem is treated similarly: one defines an arbitrary cutoff and then checks that the result is robust.
There is the additional problem that the $AdS$ geometry is a mirror for particles that try to reach infinity. That means that most null geodesic orbits never get to infinity and return back instead.
The only null geodesics that can reach infinity must have one of the angular momentum vanish (the one that corresponds to angular momentum on the $S^5$). This angular momentum counts as a mass for the particle in $AdS$ and all massive $AdS$ geodesics fail to reach infinity. This is taken care of by specializing to the correct set of geodesics.

To see this effect in the LLM coordinates, notice that the function $z$ satisfies a linear elliptic PDE with bounded boundary conditions similar to a Laplace equation. Since the region in white extends to infinity and the black region does not, for asymptotically large $r$ the value of $z$ tends to that of the white region and produces an effective angular momentum barrier repulsion unless the angular momentum of the corresponding sphere vanishes. This is how the reflection from the boundary is seen in these coordinates. Asymptotically, this is the three sphere that arises from the $S^5$, rather than the $AdS_5$. From the effective reduction to particles moving in $AdS_5$ with some mass, the mass is given essentially by the angular momentum and these trajectories cannot reach infinity. Escape only makes sense for the orbits that are effectively massless in $AdS_5$.

For a single orbit (or for a vanishingly small cell of initial conditions in phase space), the escape rate is just the inverse of the escape time: $\gamma_0=1/t_\mathrm{esc}$ (or $\gamma_0=N_0/t_\mathrm{esc}$ for $N_0$ orbits starting infinitely close to each other). However, we can also define the escape rate \emph{coarse-grained over the initial conditions} $\gamma$. In this case, we start from a finite-sized (but still small) cell with $N$ orbits with different initial conditions inside the cell. The escape rate is then given by
\be
N(t)\sim N_0e^{-\gamma t}\quad\Rightarrow\quad\gamma\equiv -\frac{\dot{N}}{N},
\ee
which of course only holds until some late time when the ensemble is almost depleted. 

It is necessary to make the cell sufficiently small so that it does not encompass very different regions of phase space (chaotic sea, stability islands, KAM tori etc). As can be seen in Fig.~\ref{figGamma_disc_ring1}, the main message is that, due to the highly complicated structure of the mixed phase space, there are several different orbit populations with distinct escape rates, which dominate over different timescales. These few escape rates repeatedly appear as we zoom into the phase space and determine the fractal structure of chaos. The correct definition of $\gamma$ also requires that we take $t\to \infty$ and take a very large ensemble because there will be fluctuations associated to individual trajectories in an ensemble.  This becomes  impractical, as it requires evolving very many  geodesics for a very long time. We have struck a balance to get an approximate measure of $\gamma$.

\begin{figure}[ht]
    \centering
    (A)\includegraphics[width=.45\linewidth]{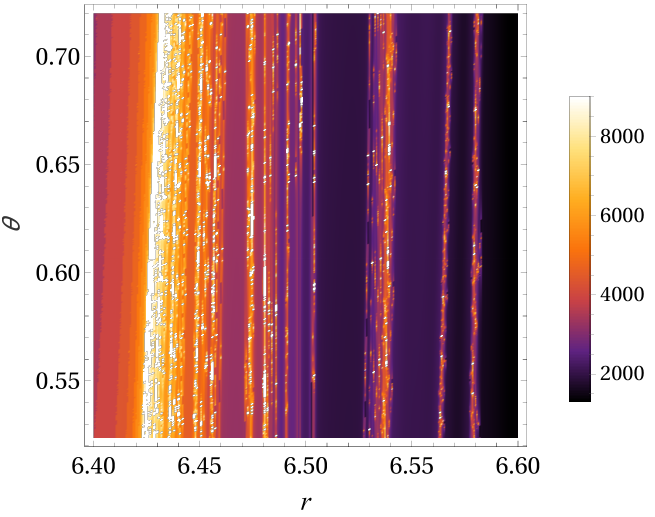}
    (B)\includegraphics[width=.45\linewidth]{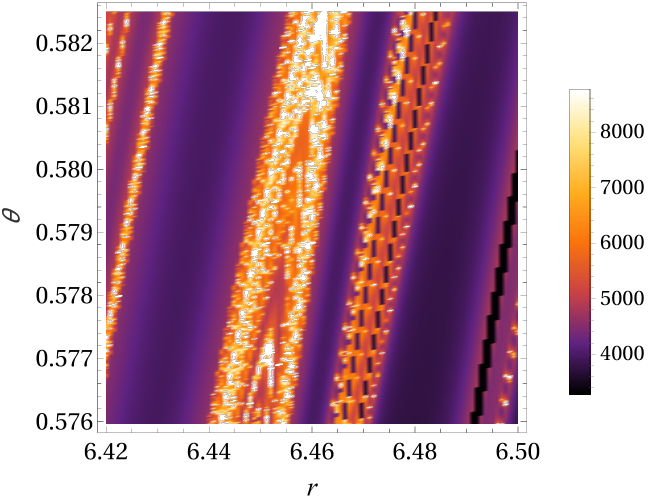}
    \caption{Escape time for a regular grid of initial conditions in $(r,\theta)$, with $P_r^{(0)}=-0.001,P_\theta^{(0)}=0.01$ and integrals of motion $(E,J_-,J_+,P_\phi)=(0.1,0.01,0.01,0.002)$. Every initial condition along $r$ and $\theta$ is color-coded for escape time. In (A) we recognize the complex and jagged boundaries of slow- and fast-escape basins, in the zoom-in shown in (B) we recognize the typical shapes of cantori and the remnants of the stability islands. The system shows a typical mixed phase space.}
    \label{figesctimedr2d}
\end{figure}

\begin{figure}[ht]
    \centering
    (A)\includegraphics[width=.28\linewidth]{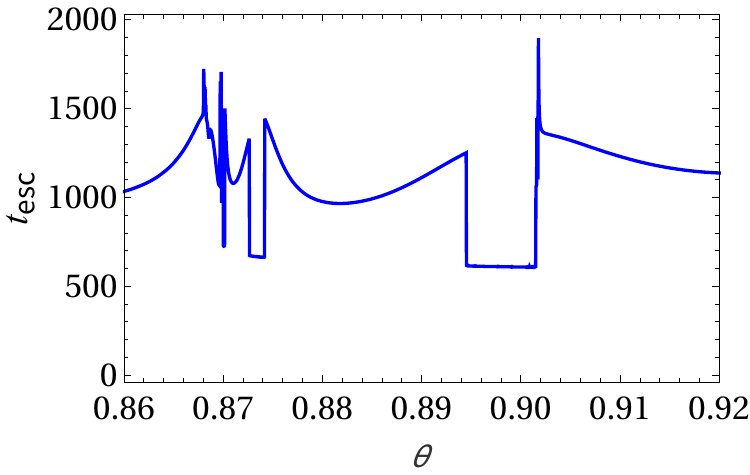}
    (B)\includegraphics[width=.28\linewidth]{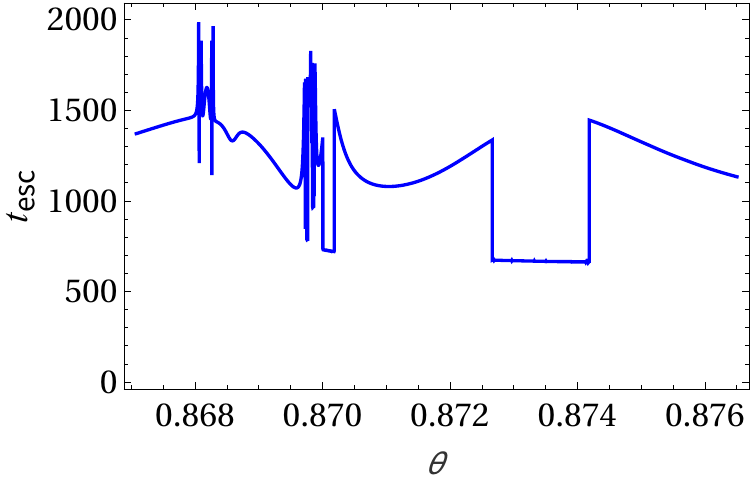}
    (C)\includegraphics[width=.28\linewidth]{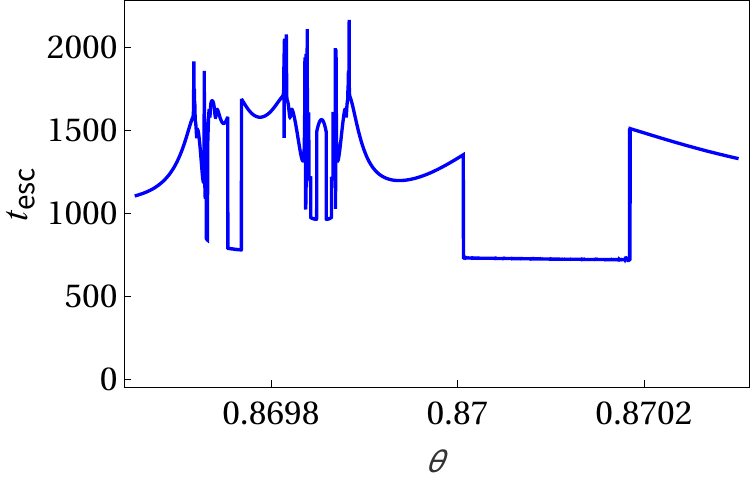}
    (D)\includegraphics[width=.28\linewidth]{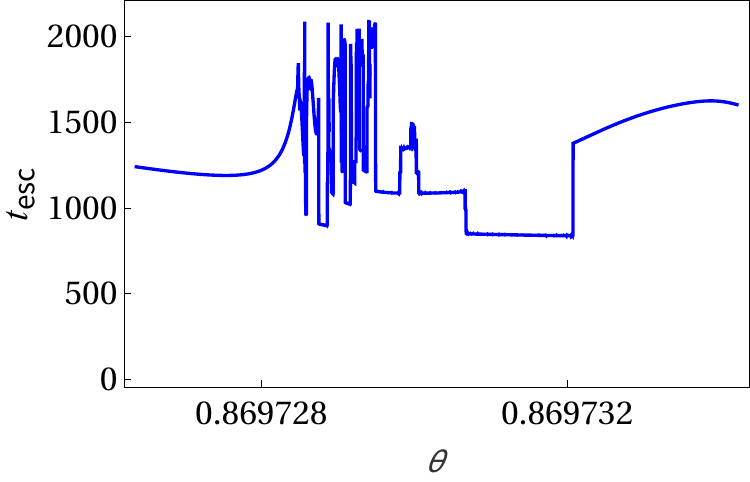}
    (E)\includegraphics[width=.28\linewidth]{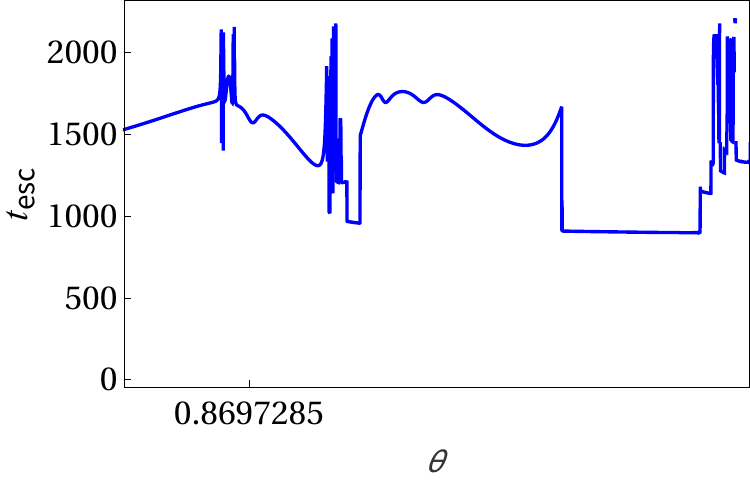}
    (F)\includegraphics[width=.28\linewidth]{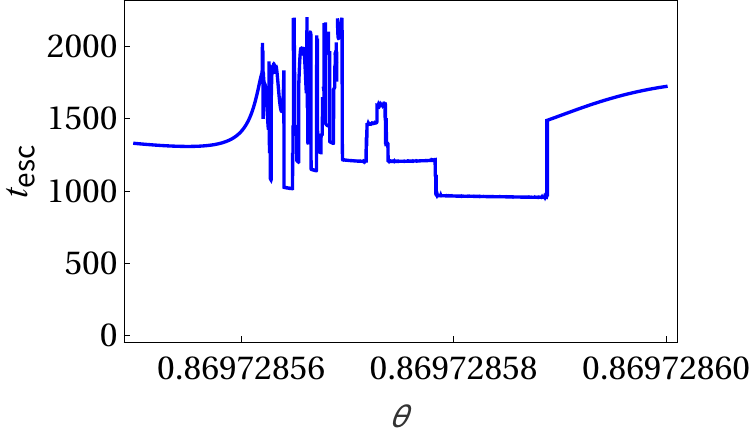}
    \caption{Escape time as a function of the angle $\theta$ at six zoom levels, showing approximately self-similar structures. The initial conditions and integrals of motion are the same as in Fig.~\ref{figesctimedr2d} (what we show here are thus 1D sections of Fig.~\ref{figesctimedr2d}).}
    \label{figesctimediskring1d}
\end{figure}

While the averaged escape rates are more appropriate for trapping and transport calculations, for the understanding of dynamics the escape times $t_\mathrm{esc}$ (or equivalently non-averaged escape rates $\gamma_0$) are more useful. In Fig.~\ref{figesctimedr2d} we can visually identify the complex structure of the phase space, in particular the high KS entropy and the remnants of invariant tori (in the zoom-in panel (B)). Linear sections of Fig.~\ref{figesctimedr2d}, shown in Fig.~\ref{figesctimediskring1d}, make the approximate self-similarity of the phase space particularly obvious. %The relation of the escape rate to their Hausdorff dimension is given in Fig.~\ref{figesctime3disk1d}. At six different scales of magnification we identify the same structures, allowing us to estimate the Hausdorff dimension:
%\be
%d_H=\frac{\log N}{\log a}=XXX.
%\ee

\begin{figure}[ht]
    \centering
    \includegraphics[width=.85\linewidth]{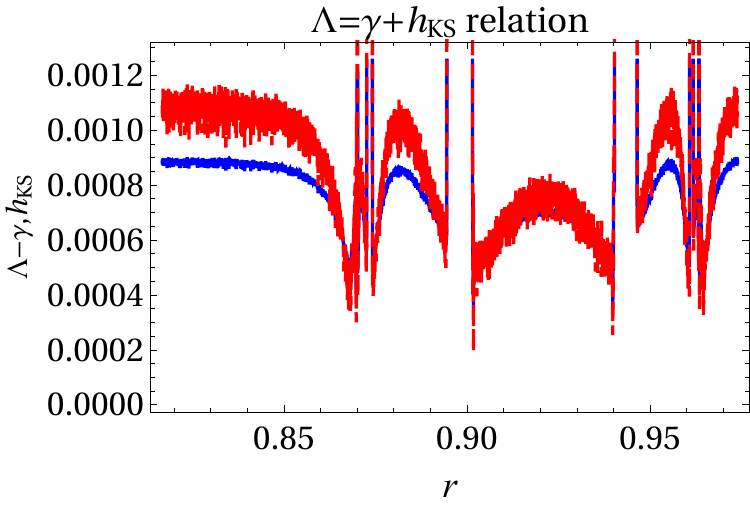}
    \caption{The numerical check of the Pesin relation (\ref{eqrelation}) between the KS entropy $h_\mathrm{KS}$, escape rate $\gamma$ and the sum of positive Lyapunov exponents $\Lambda$, for the disk+ring system. We plot the KS entropy (red) versus the difference $\Lambda-\gamma$ (blue) which is expected to equal the KS entropy. While the fluctuations are large, there is obviously a strong correlation between the two quantities. The calculations are done for the disk+ring system with the initial conditions $r^{(0)}=20,P_r^{(0)}=-0.02,\theta^{(0)}=0.275\pi\pm 0.025\pi,P_\theta^{(0)}=-0.001,\phi^{(0)}=\pi/8$ and integrals of motion $(E,J_-,J_+,P_\phi)=(0.1,0.01,0.01,0.002)$.}
    \label{figrelationdiskring}
\end{figure}

Finally, we want to check the Pesin relation \cite{Pesin:1977,Eckmann:1985,Seoane:2013} between the escape rate, the sum of positive Lyapunov exponents and the Kolmogorov-Sinai entropy:
\be 
\Lambda\equiv\sum_{\lambda_+>0}\lambda_+=h_\mathrm{KS}+\gamma.\label{eqrelation}
\ee
Besides being a litmus test of developed Hamiltonian chaos, it is also conceptually important as it provides a connection to diffusion and transport, which are naturally related to the black hole membrane paradigm. From the original derivation in \cite{Pesin:1977}, the ensemble-averaged escape rate $\gamma$ is the right one to consider as the derivation starts from a measure of non-escaping orbits in phase space. The numerical check of the $\Lambda-h_\mathrm{KS}-\gamma$ relation for the disk+ring configuration is found in Fig.~\ref{figrelationdiskring}.

%\begin{figure}[ht]
 %   \centering
  %  \includegraphics[width=.85\linewidth]{hKS_disc_ring2.pdf}
   % \caption{Decay of the number of occupied cells in the phase space as a function of proper time $N_o(\tau)$ for a set of null geodesic orbits in the LLM geometry sourced by disk + ring pattern in a plane. Integrals of motion for this set of orbits are $\{ L = 0.001, \widetilde L = 0.001, E=0.02 \}$. Kolmogorov-Sinai entropy per unit time is obtained from a characteristic $N_o(\tau) \sim e^{- h_{KS} \tau}$ behavior: $h_{KS} \approx 0.0000468614$.}
    %\label{fighKS_disc_ring}
%\end{figure}

%\mc{comment on diffusion}

\subsection{Multi-disk geometries}

\begin{figure}[ht]
    \centering
    (A)\includegraphics[width=.45\linewidth]{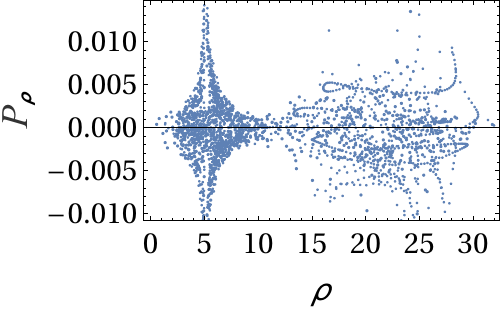}
    (B)\includegraphics[width=.45\linewidth]{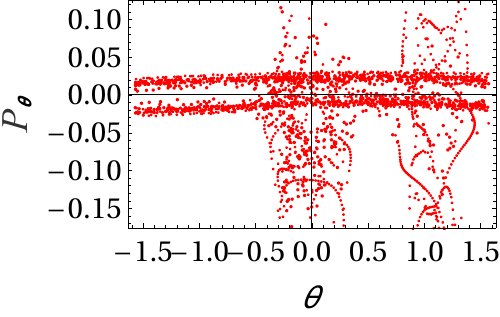}
    \caption{Poincare section $P_{\xi} = 0$ in $P_\rho-\rho$ and $P_{\theta}-\theta$ planes for a set of null geodesic orbits in the LLM geometry sourced by the 3-disk pattern in a plane. The chaotic sea is more dominant than for the disk+ring case. Integrals of motion for this set of orbits are $\{ J_- = 0.001, J_+ = 0.001, E=0.02 \}$.}
    \label{figPS_three_disc}
\end{figure}

\begin{figure}[ht]
    \centering
    (A)\includegraphics[width=.29\linewidth]{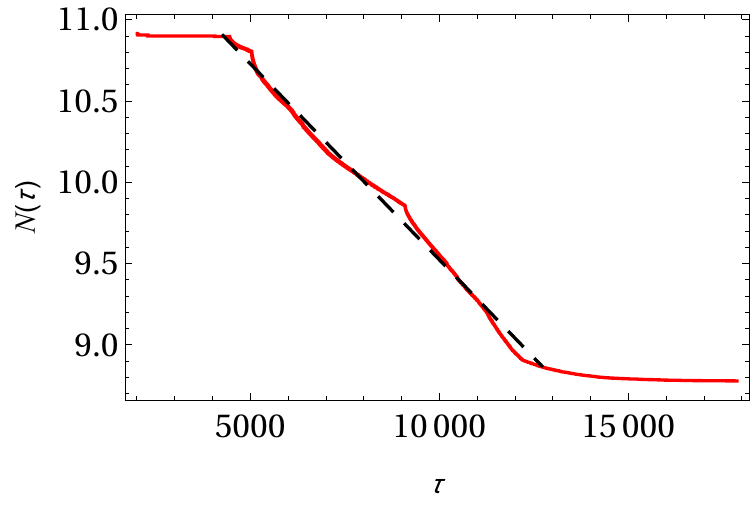}
    (B)\includegraphics[width=.29\linewidth]{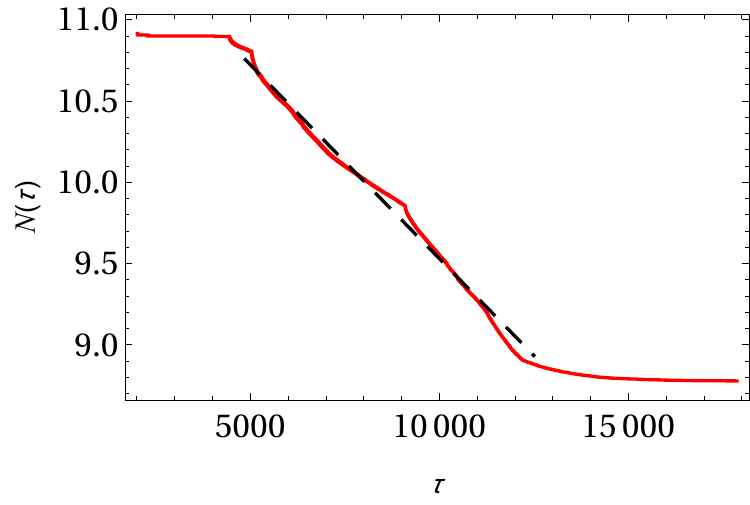}
    (C)\includegraphics[width=.29\linewidth]{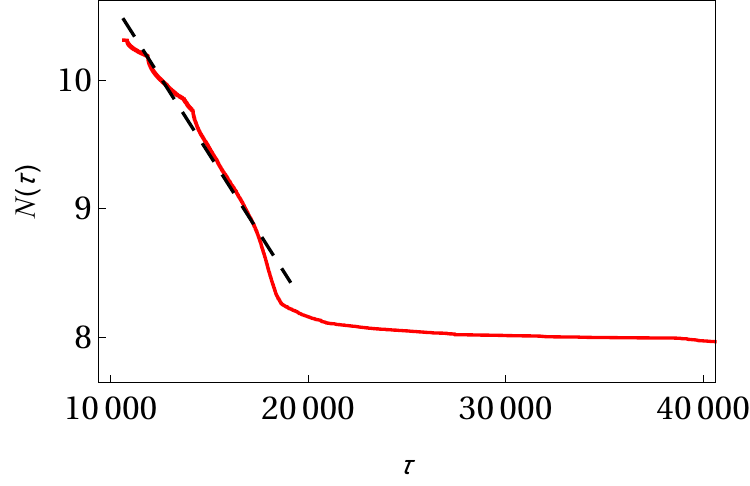}
    \caption{Decay of the number of trapped orbits as a function of proper time $N(\tau)$ (red full lines) for the 3-disk geometry, for three cells of initial conditions of decreasing size, given by $\xi^{(0)}=4.7, P_\xi^{(0)}=-0.01$, $\phi^{(0)}=\pi/8,P_\phi^{(0)}=0.002$, $P_r^{(0)}=0.001$ and $r^{(0)}$ from the interval $r^{(00)}\pm\Delta r$, with $\Delta r=0.3,1.0\times 10^{-9},5.-\times 10^{-13}$ (A to C respectively). The black dashed lines yield the best linear fit, the slope defining the escape rate $\gamma$ from $N(\tau)\sim\exp(-\gamma\tau)$. The 3-disk case generally exhibits stronger chaos, yielding a single, rather universal escape rate $\gamma\approx 0.0023\pm 0.0002$.}
    \label{figGamma_3disks}
\end{figure}

Even though the geodesic dynamics is now truly 3-dimensional, with no extra integrals of motion, it is still convenient to transform the Cartesian coordinates in the LLM plane to the polar ones ($x=\rho\cos\phi,~y=\rho\sin\phi$), while keeping the ``altitude'' $\xi$ (corresponding essentially to cylindrical coordinates). The Hamiltonian (\ref{eqHa}) for a null geodesic in these coordinates reads:
\begin{equation}
    2 h^2 ~\mathcal H = P_{\xi}^2 + (P_\rho + E V_\rho)^2 + \frac{\left( P_{\phi} + E V_{\phi} \right)^2}{\rho^2} - h^4 \left( E^2 - \frac{2J_-^2}{1 - 2z} -\frac{2J_+^2}{1 + 2z} \right).
\end{equation}
We can again look at the Poincare surface of section defined by $P_\xi=0$. The corresponding sections in $P_\rho-\rho$ and $P_\theta-\theta$ planes are shown in Fig.~\ref{figPS_three_disc}. In comparison with the disk+ring case it is more chaotic, as expected given the fact that we have one more degree of freedom (i.e. one integral of motion less). Similar conclusions can be gained from the more uniform and universal behavior of the escape rate, shown in Fig.~\ref{figGamma_3disks} -- now there is only one rate, i.e. only one chaotic component, as the surviving invariant tori do not form barriers in phase space.

%\begin{figure}[ht]
%    \centering
%    (A)\includegraphics[width=.45\linewidth]{Gamma_three_discs.pdf}
%    (B)\includegraphics[width=.45\linewidth]{Gamma_three_discs_Trapping.pdf}
%    \caption{Escape rate along $r$-direction (A) and along the $\xi$-direction (B), for a set of null geodesic orbits in the LLM geometry sourced by the 3-disk pattern in a plane. Integrals of motion for this are $L = 0.001, \widetilde L = 0.001, E=0.02$ for both panels, and the initial conditions are $\xi^{(0)} = 10, P_{\xi}^{(0)} = -0.01$ \mc{for (A) too? You didn't give anything for (A). What is the cell size?}. In (B) the orbits evolve towards the LLM plane and get trapped near it for a while before they escape \mc{trapping?}. The escape rates are $\bar{\gamma}=1.17 \times 10^{-4}$ (A) and $\bar{\gamma}=4.36\times 10^{-4}$ (B).
%    \vd{ plot vs $t(\tau)$ not $\tau$}}
%    \label{figGamma_three_disc}
%\end{figure}

The phase space is again mixed, with stability islands and cantori dispersed in the chaotic sea, but we expect this to have less influence on averaging as the phase space dimension is now 6 instead of 5, thus the chaotic orbits have a lower chance to bump into quasiregular structures.

\section{Null geodesics in grayscale LLM geometries and ensemble averaging}\label{secgray}

\subsection{Grayscale LLM geometries}

Grayscale LLM geometries are best understood as coarse-grained LLM solutions \cite{Balasubramanian:2018yjq}. In the matrix model \cite{Berenstein:2004kk}, they are obtained from ``smoothed'' Young tableaux which are only defined up to an order $O(1)$ fluctuation in the number of squares in each row (i.e. where we cannot detect adding/removing $O(1)$ squares); in the supergravity picture they correspond to ``smoothed'' boundary conditions where the $z$-functions from Eq.~(\ref{eqllmz}) are allowed to have any value of gray between $-1/2$ and $1/2$:
\be
z(\xi\to 0)=b,~~-1/2\leq b\leq 1/2.\label{eqgraybc}
\ee
This restriction is forced on us by positivity of the metric of the three spheres away from the singularity (one can relate the fact that chronology should be protected to unitarity of the free fermion model and its relation to the Fermi exclusion principle  \cite{Caldarelli:2004mz}). 

Looking at our Hamiltonian \eqref{eqHaconf}, we notice that the effective potential is given by
\begin{equation}\label{eq:effectivepot}
V_\mathrm{eff}= - h^4 \left( E^2 - \frac{2J_-^2}{1 - 2z} - \frac{2J_+^2}{1 + 2z} \right) ,
\end{equation}
and notice that since $z$ is finite as $\xi \to 0$, none of the terms with angular momentum are more divergent near the grayscale LLM plane region than the term that also includes $E$. The divergence in the potential in this case comes from $h^{-2}\to 0$ as $\xi\to 0$, which does not have a fixed sign.
In black and white setups, $h$ is generically finite near $\xi\to 0$ and one of the two terms with angular momentum produces a repulsive force that overwhelms the term with the energy $E$, so in black and white patterns trajectories are generically repelled when they get sufficiently close to the LLM plane. 
To make sense of the fact that a grayscale pattern should be thought of as averaging over geometries, this repulsion must be happening on distances that are small relative to the coarse-graining scale. 

For a single disk, this gives the solution for $z$ which is exactly the same as for the black disk on white background (Eq.~\ref{zsinglesol}), but multiplied by $g$: $z(\xi,x,y,R)\mapsto gz(\xi,x,y,R)$. The generalization for multiple disks is obtained analogously, multiplying each disk solution by the appropriate flux. As we approach the LLM plane $\xi =0$, the most diverging metric coefficients in any solution with a gray disk scale as $\sim h^2 \sim \sqrt{ 1/4 - z^2 } /\xi$. Therefore, one should expect that the scalar curvature in the LLM plane diverges as violently as $\xi^{-3}$.

On the other hand, there is still no horizon. We have thus produced a singularity but not a black hole. However, it is a good singularity in the Gubser scheme \cite{Gubser:2000nd}, and one should in principle be able to construct a black hole out of it. Such an object begs the question that is now of utmost importance in quantum gravity: \emph{is the coarse-graining/averaging a fundamental one, i.e. averaging over an ensemble, or is it an effective averaging due to finite resolution, i.e. averaging over small details of a given fixed LLM geometry?} Our central idea is to answer this question in detail for geodesics and their dynamical properties.

The picture that emerges is that when we use a grayscale averaging, the singularity becomes effectively attractive and more similar to a black hole with a tiny horizon, where we would expect similar phenomena.
This can also be explained as follows. In the metric \eqref{eqllmg}, the function $h^{-2}$ is $g_{tt}$, the warp factor of the time direction. If $h^{-2}\to 0$, there is a region near the singularity where time is warped very much and it produces a gravitation potential barrier to get out. This warping is typical of near horizon regions, where the horizon is usually characterized by $g_{tt}=0$ in more conventional setups. 

\subsection{Geodesics in grayscale geometries}

Although no new integrals of motion appear in grayscale, and the system is still nonintegrable, the dynamics is much simpler now. We now have two ``attractors'': the singularity in addition to infinity, as follows from looking at the effective potential \eqref{eq:effectivepot} \footnote{Of course, there is no attractor in a Hamiltonian system. However, in Hamiltonian scattering, infinity can be thought of as an attractor (for unbounded orbits), and in our case the singularity is also attractor-like in the sense that an open set of orbits near it cannot get away from it.}. Naively, one might think that this would lead to even more complex behavior, but it turns out that most orbits are short and quickly end either at infinity or in the singularity. 

\begin{figure}[ht]
    \centering
    (A)\includegraphics[width=.45\linewidth]{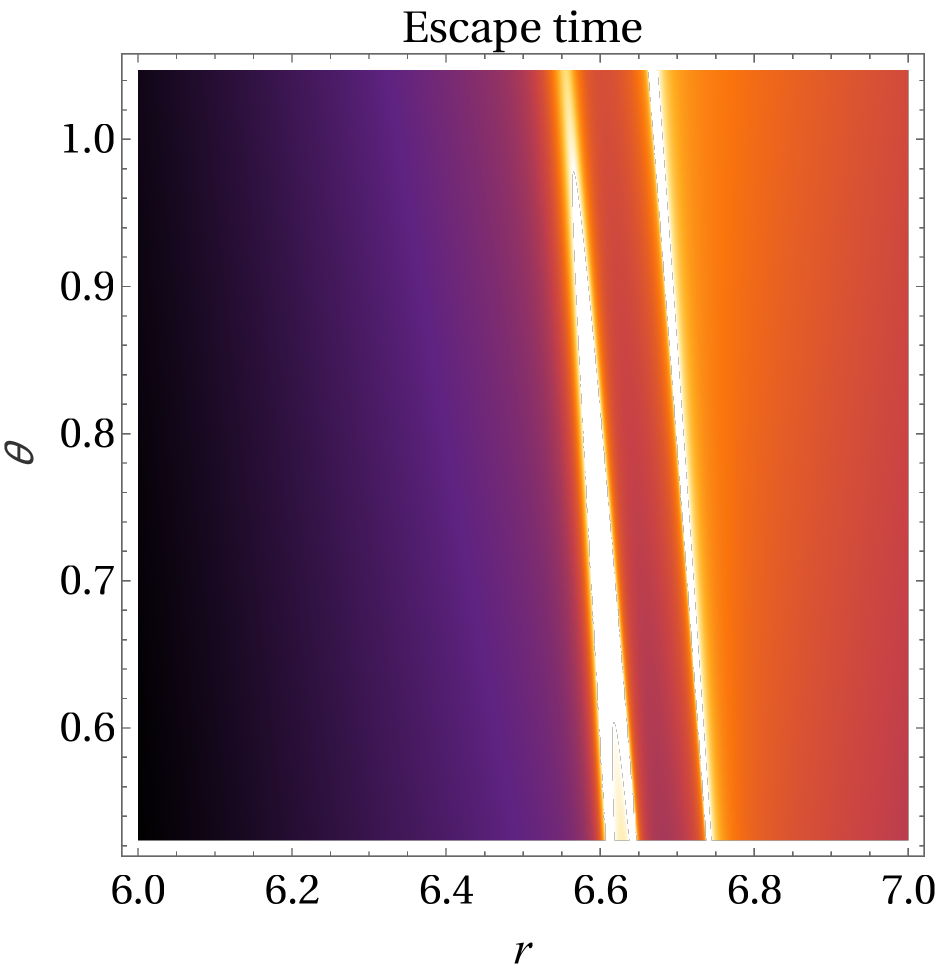}
    (B)\includegraphics[width=.45\linewidth]{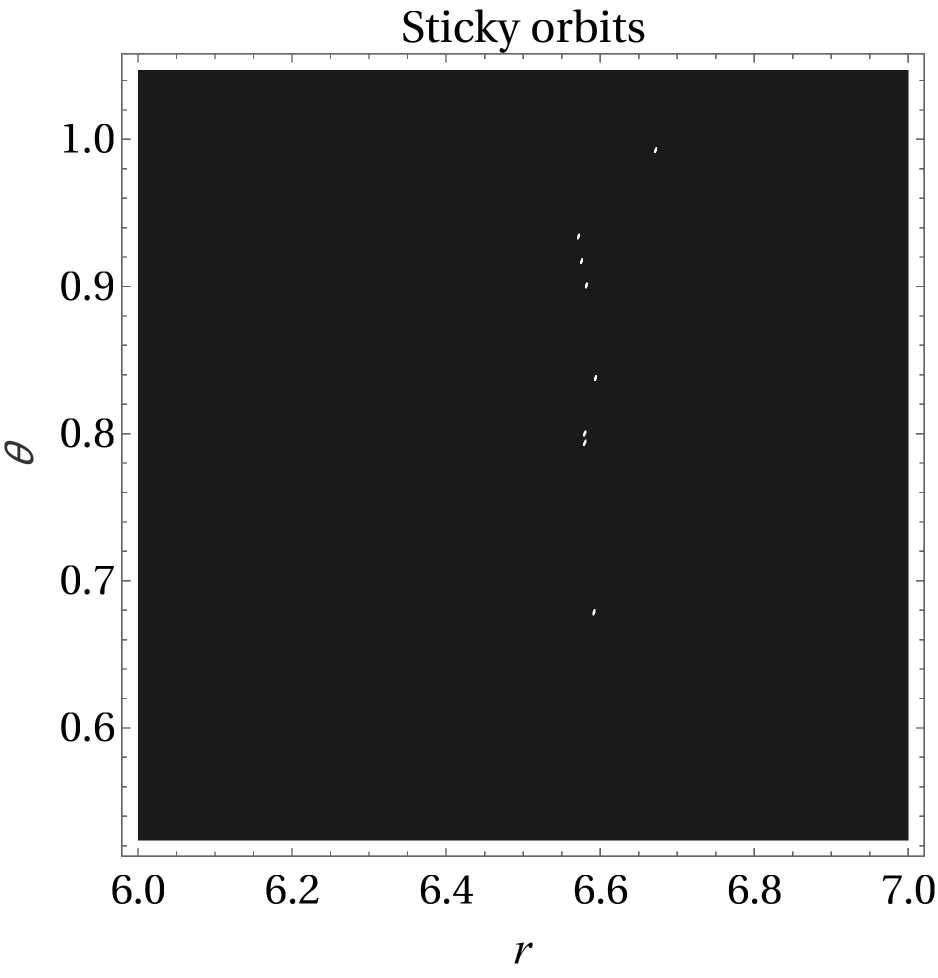}
    \caption{Escape time for a regular grid of initial conditions in $(r,\theta)$, with $P_\theta^{(0)}=-0.01,P_r^{(0)}=0.001$ and integrals of motion $(E,J_-,J_+)=(0.1,0.01,0.01)$. In (A) every initial condition is color-coded for escape time -- the dependence on initial conditions is smooth and no zooming is necessary as there are no sticky islands or similar complex structures. In (B) we indicate the sticky orbits ($t_\mathrm{esc}/t_0>50000$): there are only very few such orbits (and even these might be numerical artifacts) and we can conclude there is neither stickyness nor developed chaos. The color code in (A) is the same as in Fig.~\ref{figGamma_disc_ring1}.}
    \label{figGamma_disc_ringgray}
\end{figure}

\begin{figure}[ht]
    \centering
    (A)\includegraphics[width=.45\linewidth]{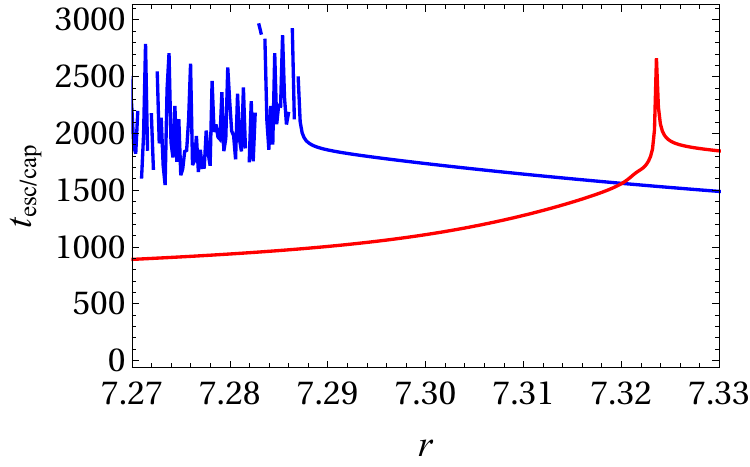}
    (B)\includegraphics[width=.45\linewidth]{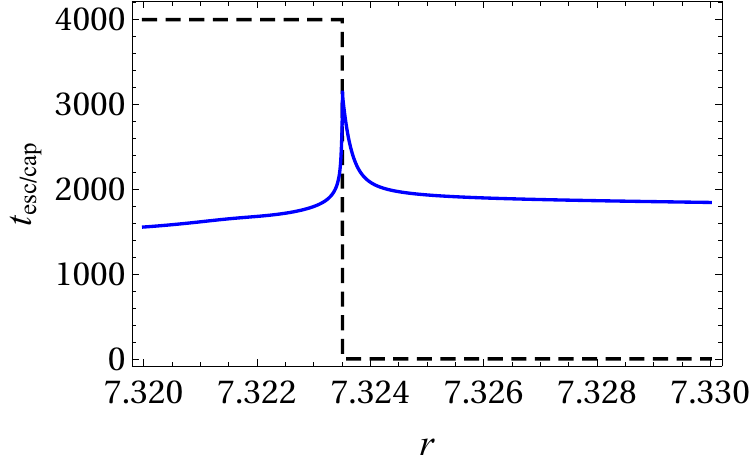}
    \caption{(A) Escape/capture time as a function of $r$ for grayscale disk+ring geometry (red) and for a specimen from an ensemble of $N=31$ black and white rings (blue). Only the latter shows a complex dependence of the escape time on initial condition, indicative of chaotic scattering: everything is washed out in grayscale. In (B) we show that the critical point in ther grayscale escape/capture time, where the time seemingly diverges, corresponds to the bifurcation between escaping ($r<r_0$) and captured ($r>r_0$) orbits, with $r_0\approx 7.3235$.}
    \label{figesctimediskringgray}
\end{figure}

Fig.~\ref{figGamma_disc_ringgray} presents the variation of escape time with initial conditions analogously to Figs.~\ref{figGamma_disc_ring1} and \ref{figGamma_3disks}. Now the dependence is smooth, without complex structures, and panel (B) additionally shows that sticky orbits are very few and far apart. In Fig.~\ref{figesctimediskringgray}(A) the one-dimensional cut along the $r$-axis explicitly compares the grayscale geodesics (red) with the black and white ones (blue), again showing that the strongly chaotic region vanishes. In Fig.~\ref{figesctimediskringgray}(B) we delineate the unstable periodic orbit dividing the escaping orbits ($r>r_0$) from those which fall into the singularity ($r<r_0$). This is analogous to the black hole's photon ring, and might be interestingly related to the quasinormal mode spectrum; we postpone this line of thinking for further work. Here, our main point is the simplicity of dynamics: although nonintegrable, it comes closer to the regular behavior of geodesics on most black hole backgrounds.

\subsection{Grayscale vs. averaging over black-and-white}

%%%%%%%%%%%%%%%%%%%%%%%%%%%%%%%%% This is new

\subsubsection{On the origin of grayscale geometries}

So far we have considered grayscale geometries, but we have not stated their origins. There are a number of approaches that can lead to effectively averaged solutions at the supergravity level. One approach is to start from the free fermion description of half-BPS states \cite{Berenstein:2004kk}: any reasonable wave function for the fermions (i.e., a normalizable state with bounded energies, etc) is a state that might be describable by supergravity solutions that we have considered above. For example, we can consider a fractional quantum Hall state wave function. It is an example of a physically reasonable state, but not a ground state for the free fermion Hamiltonian. Such a wave function is a ground state wave function of a different type of incompressible fluid and one can check that it has constant density on the occupied region of the fermion phase space. It is not easy to describe it in terms of the Young tableaux basis that spans the Hilbert space, but it is one possible wave function of the system that some clever experiment might have prepared for us. In that sense, it is possible to imagine that we can generate a single microstate with different grayscale factors in different regions, where the precise given wave function that generates such a pattern might be very complicated.

A slightly different approach is to consider a non-trivial set of  concentric ring geometries with excitations on top of them that are entangled between different edges of the geometry. 
In \cite{Berenstein:2017abm} it was shown how to explicitly write down such excitations; one can then build an approximate squeezed state between different such excitations that in expectation values produces a grayscale pattern locally. Concerning the local degrees of freedom near the corresponding edge, since they are entangled with other degrees of freedom that are far away, the local physics can be thought of as a generalized ensemble of geometries with thermal-looking modes, with a different temperature for each mode. This would be a version of a generalized Gaussian ensemble for the local degrees of freedom.

Basically, we can justify building a locally preferred grayscale pattern by averaging over some ensemble of choice that can be used to represent these options. We can then ask if the physics of the probes we are considering (the lightlike geodesics) are sensitive to the details. That is, can we distinguish a fundamental grayscale state from a state that is gray just because a very large class of wave functions locally can be thought of as a classical statistical ensemble of geometries near the region we want to study? At the level of our present analysis, the answer is no: the geodesics do not see the wave function, i.e. the microstate, and are solely determined by the geometry, which is the same no matter what is the origin of the coarse-graining. Our goal is then to build a simple model as an example of this idea.

%That is, can we distinguish a grayscale that came from a state that is more fundamentally thought of as a grayscale state to one that is 

%%%%%%%%%%%%%%%%%%%%%%%%%%%%%%%%%%%%% End new

\subsubsection{Averaging in a toy model}

We will again consider the disk+ring geometry but now the area between the disk and the ring is grayscale, with a value between $-1/2$ and $1/2$. The equations for the metric remain the same as (\ref{eqllmg}-\ref{eqllmz}) except that the boundary conditions (\ref{eqllmbc}) are replaced by (\ref{eqgraybc}). This solution can be understood as the coarse-grained sequence of black and white rings with separation less than $\sqrt{\hbar}$, the coarse-graining scale. In order to understand how the averaging works, we will compare the grayscale geometry with a coarse-grained sequence of black and white rings of finite separation, which nevertheless should capture the basic workings of the averaging when the total number of rings is large.

This works at values of $\xi$,  where $\xi$ is strictly non-zero and fixed, as we take the concentric rings to be smaller and smaller. Only then, we are allowed to take the limit $\xi \to 0$. Basically, the coarse-graining procedure requires taking a particular order of limits. 
Notice that in the $\xi$ variables, the warp factor of the grayscale geometry satisfies $g_{tt}\to 0$ with a single zero at $\xi=0$, typical of black hole horizons. The metric, restricted to the $t,\xi$ coordinates near this region looks like $-\xi dt^2+d\xi^2/\xi$, which is a metric of Rindler space in two dimensions and starts looking like a black hole throat near the singularity. Orbits that fall to the $\xi\to 0$ region take a long time to get out because of the warp factor. Essentially, when we integrate the time variable in the geometry relative to $\xi$ we find that the time scales as  $t\simeq -\log(\xi)$. This is different than the effective time of the Hamiltonian system we solve, which is an affine parameter of the null geodesic. The region 
near the singularity is very similar to a near horizon geometry of a black hole and it traps orbits for a long time. 

\begin{figure}[ht]
    \centering
    (A)\includegraphics[width=.45\linewidth]{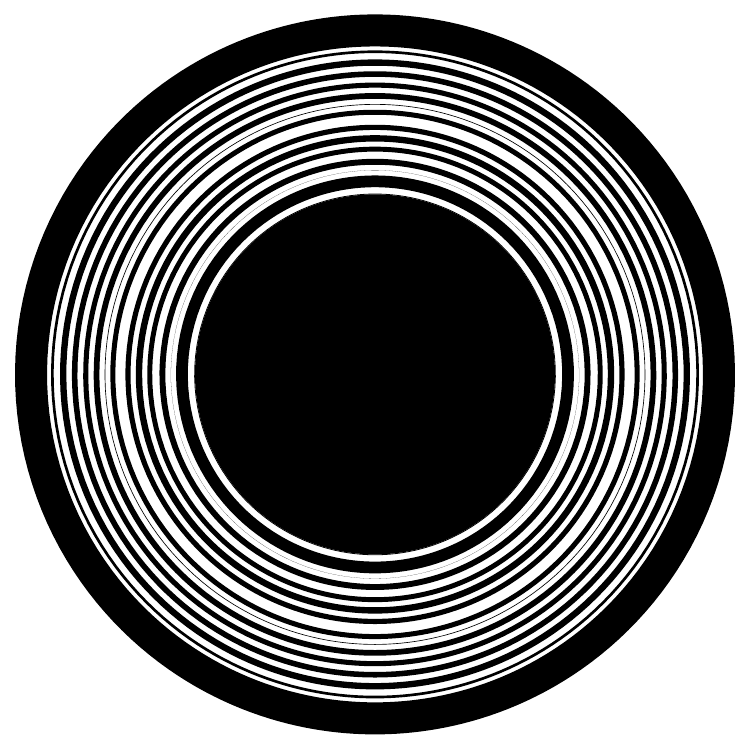}
    (B)\includegraphics[width=.45\linewidth]{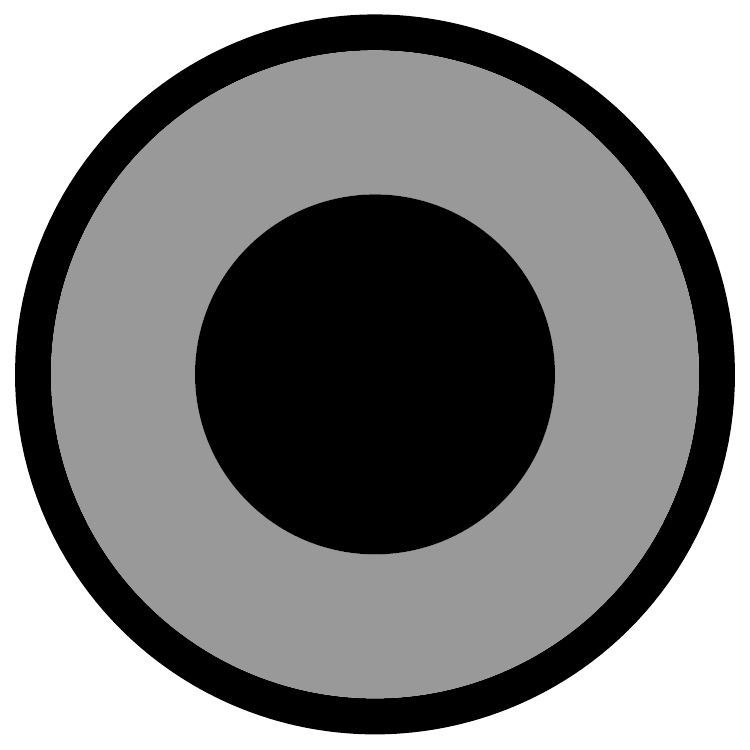}
    \caption{Geometry with a disk and $N$ thin black rings (A) should be well-approximated by a coarse-grained solution with a single gray ring (B). Our main puzzle, considered in this section, is if this averaging carries over to the geodesic fluctuations.}
    \label{figgrayrings}
\end{figure}

Going back to the averaging procedure,  consider a specific model of $2N+1$ concentric rings, where the black disk of radius $R_1$ is followed by a white ring extending to $R_2$, followed by a black ring to $R_3$ and so on, ending with a black ring between $R_{2N}$ and $R_{2N+1}$ (Fig.~\ref{figgrayrings}(A)).\footnote{It is easy to see that the total number of disk and ring boundaries must be odd, hence $2N+1$.} For $R_i-R_{i-1}$ small ($2\leq i\leq 2N$), a grayscale geometry with a gray ring extending from $R_1$ to $R_{2N}$ (Fig.~\ref{figgrayrings}(B)) should be a good approximation. In order to approximate the gray area, we average over $2N-2$ numbers: $R_i$ with $2\leq i\leq 2N-1$; $R_1$, $R_{2N}$ and $R_{2N+1}$ are fixed. The probability density $P(R_2,\ldots R_{2N-2})$ must be normalized:
\be
\int dR_2\ldots\int dR_{2N-2}P(R_2,\ldots R_{2N-2})=1\label{pdist}
\ee
and conserve the total flux through the LLM plane. In the language of the matrix model this is the Luttinger theorem (the total number density is conserved). For our concentric circles model this means (it is understood that $R_0\equiv 0$)
\be
\sum_{i=1}^N\left(R_{2i+1}^2-R_{2i}^2\right)\equiv\frac{\mathcal{A}}{\pi}=R_1^2+g\left(R_2^2-R_1^2\right)+R_3^2-R_2^2.\label{grayconstraint}
\ee
%%%%%%%%%%%%%%%%%%%%%%%%%%%% New stuff

As we have already described, we are justified in using such an ensemble because local classical features can appear to be in a mixed state in the quantum mechanical wave function either because we are considering a true \emph{ensemble} of geometries with similar features, or because local features can \emph{appear} to be mixed since they are entangled with degrees of freedom that are far away. Many such wave functions could lead to the same local ensemble, which is justified by stating that a density matrix in the particular Young diagram basis is built to be (locally) diagonal in that basis as a member of a general class of ensembles with similar properties. 

We can also consider a slightly different point of view: take a superposition of different classical ring geometries as a type of Schr\"odinger  cat state. Since classically they are different, any sufficiently reasonable classical measurement would be diagonal in this basis and would lead to a distribution like the one determining $P$ in (\ref{pdist}), which would be sensitive to the amplitudes of the superposition, but not to the phases of individual states.

%%%%%%%%%%%%%%%%%%%%%%%%%%%%%%%%

Now the average of any observable $O(x(\tau))$ which is a function of a geodesic $x(\tau)$ is obtained as
\be
\langle O\left(x\left(\tau\right)\right)\rangle=\prod_{i=2}^{2N-1}dR_iP(R_2,\ldots R_{2N-2})O(x(\tau)),
\ee
where it is understood that the geodesic $x(\tau)$ implicitly depends on the background and thus on $R_i$. Ideally, one would like to directly calculate the average of a geodesic length or deflection angle, which are relevant for the calculation of correlation functions in the eikonal approximation. We instead do a simpler and more tractable task: averaging the effective potential felt by the geodesic. This should suffice to give an idea of whether the averaging goes through or not.

%Analogously to the procedure of the backgrounds themselves, we may average any operator. Specifically, for a geodesic $x^\mu(\tau)$ we have
%\be
%\bar{x}^\mu(\tau)=\int dR_2\ldots\int dR_{2N-2}P(R_2,\ldots R_{2N-2})x^\mu(\tau;\lbrace R_2,\ldots R_{2N-2}\rbrace),
%\ee
%where $x^\mu(\tau;\lbrace R_2,\ldots R_{2N-2}\rbrace)$ is the geodesic for given fixed black and white configuration. We first solve (\ref{grayconstraint}) for $R_2$ and then assume a uniform distribution of $R_3\ldots R_{2N-2}$. 

%On the other hand, we can formulate the averaging over orbits with different initial conditions $\mathbf{r}_0$, distributed according to some $\mathcal{P}(\mathbf{r}_0)$ (all orbits now move in the same geometry):
%\be
%\langle x^\mu(\tau)\rangle=\int\mathbf{r}_0\mathcal{P}(\mathbf{r}_0)x^\mu(\tau;\mathbf{r}_0).
%\ee
%These two averages correspond to the averaging over theories (as in JT gravity) and the averaging over XXX.

\subsection{Near-plane effective potential}

\subsubsection{Black and white vs. gray potential}

Even this problem, the analysis of the effective potential, is overall analytically intractable. But the dynamics is largely dominated by bounces off the LLM plane, and in this regime the effective potential is dominated by the discontinuity at the black-white boundary. This boundary in turn corresponds to the $AdS$ boundary of a given disk ($AdS$) region, and the effective potential acquires the familiar inverse-square form.

We can show this explicitly by expanding the effective Hamiltonian (\ref{eqHa}) near the LLM plane, i.e. around $\theta=0$ or $\xi=0$. Expanding near $\theta=0$ and the black-white boundary $(\theta=0,r=R_i)$ and putting $P_r=P_\xi=0$, we  get
\be
%V_\mathrm{BW;eff}(\theta,r)=\frac{J_+^2+J_-^2+\left(J_-^2-J_+^2\right)\mathrm{sgn}(r^2-R_i^2)}{2r^2\theta^2}+O(\theta^0).\label{veffbwtheta}
V_\mathrm{BW;eff}(\theta,r)=\frac{J_-^2\Theta(\rho-R_i)+J_+^2\Theta(R_i-\rho)}{2r^2\theta^2}+O(\theta^0).\label{veffbwtheta}
\ee
Doing the same for a gray disk of intensity $g$ with the appropriate metric, we get
\be
V_\mathrm{gray;eff}(\theta,r)=\frac{J_+^2+J_-^2+g\left(J_-^2-J_+^2\right)\mathrm{sgn}(r^2-R_i^2)-E^2\frac{1-g^2}{2}}{2r^2\theta^2}+O(\theta^0).\label{veffgraytheta}
\ee
In the cylindrical $\xi$ coordinate the expressions are analogous:
\be
V_\mathrm{BW;eff}(\xi,\rho)=\frac{J_-^2\Theta(\rho-R_i)+J_+^2\Theta(R_i-\rho)}{\xi^2}\label{veffbw}
\ee
\be
V_\mathrm{gray;eff}(\xi,\rho)=\frac{-(\frac{E}{2})^2\left(1-g^2\right)+\frac{J_-^2+J_+^2}{2}+\frac{g}{2}\left(J_-^2-J_+^2\right)\mathrm{sgn}(\rho-R_i)}{\xi^2}\label{veffgray}
\ee
Notice the discontinuity at $r=R_i$ which persists also in grayscale. The crucial difference between the black-and-white and gray cases is that, for some interval of $g$ values, the grayscale geometry develops a potential well, corresponding to the fall into the singularity. The black and white potential is always repelling, i.e. positive. The main reason is the behavior of $h$. For example, in regular LLM geometries the function $h$ remains finite as $\xi \to 0$ (see for example \cite{Berenstein:2020jen}). That is, only one sphere radius shrinks. In the grayscale setup $h$ blows up because $z\neq \pm 1/2$. Then, even though this counts as an angular momentum barrier, the contribution of $E$ is of the opposite sign than the contributions from $J_\pm$ and this can produce a strongly attractive potential to the singularity if $E$ is large enough.
%\begin{enumerate}
%\item For $g<g_1$ (with $g_1$ depending on the parameters), the grayscale develops a potential well, corresponding to the fall into the singularity. The black and white potential is always repelling, i.e. positive.
%\item For $g<g_2$ (again with nonuniversal $g_2$), the grayscale potential is negative everywhere, with no repelling behavior at all. 
%\end{enumerate}

\begin{figure}[ht]
    \centering
    (A)\includegraphics[width=.45\linewidth]{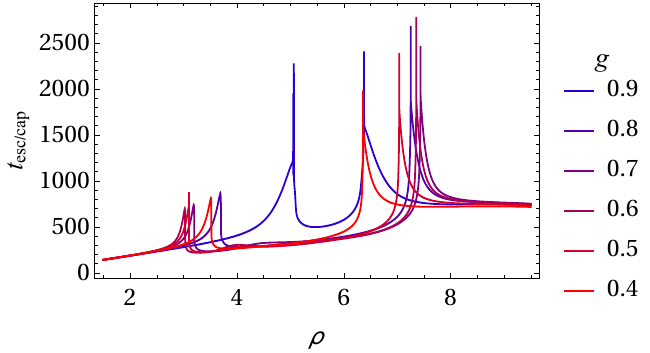}
    (B)\includegraphics[width=.45\linewidth]{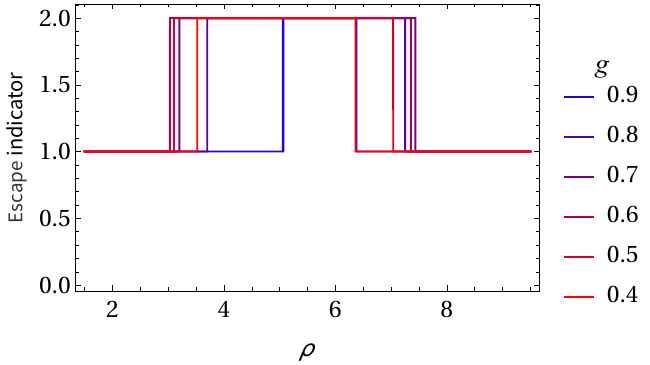}
    \caption{(A) Escape/capture time as a function of $r$ for six grayscale geometries with different fluxes, with initial conditions $\rho^{(0)}=20,P_\rho^{(0)}=0.01$ and integrals of motion $(E,J_-,J_+)=(0.1,0.01,0.01)$: the time is a smooth function of $\rho$ as already found. (B) Escape indicator for the same setup as in (A): the geodesics escape for 1 or end up in the singularity for 2. For $g<g_-\approx 0.30$ and $g>g_+\approx 0.90$ all orbits at this energy would end up escaping, because the effective potential becomes strictly positive (i.e., repulsive).}
    \label{figesctimediskringgray1}
\end{figure}

This strongly attractive potential means that many orbits cannot wander for a long time but have a simple structure, falling straight into the singularity. That in turn explains the simpler structure of the geodesics and the lack of strongly chaotic orbits. The quantitative criterion for attraction is simply that the effective potential $C/\xi^2$ has $C<0$. Equating the coeffficient of Eq.~(\ref{veffgray}) to zero we get
\be
g_\pm=\frac{\vert J_-^2-J_+^2\vert\pm\sqrt{E^4-2E^2\left(J_-^2+J_+^2\right)+\left(J_-^2-J_+^2\right)^2}}{E^2}.
\ee
The $g_+$ and $g_-$ solutions correspond to semi-white (positive) and semi-black (negative) values, respectively; for the gray nuances $g_-<g<g_+$ the geodesic falls into the center. These analytical findings are confirmed in Fig.~\ref{figesctimediskringgray1}: precisely between the "semi-white" critical value and the ''semi-black'' critical value the geodesics fall into the LLM plane. 

%\begin{figure}[ht]
%    \centering
%    \includegraphics[width=.45\linewidth]{DiskRingEscapeCrit.pdf}
%    \caption{The minimal initial condition $\rho_0$ for which the geodesic is captured as a function of the gray valeur $g$: for $g<g_1$ with $g_1\approx 0.12$ there is no capture. The dashed line denotes the minimal $\rho_0$ for escapes in black and white geometry XXX.}
%    \label{figesctimediskringgray2}
%\end{figure}

In conclusion, the naked singularity acts as a strong attractor whose domain of attraction is mainly determined by energy. As a result, most orbits of given energy either fall in or escape, and the dynamics is much simpler and more regular compared to the black and white case. Furthermore in the potential $C/\xi^2$ it takes a finite amount of time in our Hamiltonian system to reach $\xi=0$. Remember that this is an affine time of the null geodesic and not the time $t$ elapsed in the LLM coordinates that describe the infall. That external time which could be measured by an external observer is actually large.

Notice that a hint of this behavior can be gleaned from 
figures \ref{figbackgnddiskring} and \ref{figbackgnd3disks}. There, the null geodesics seem to want to go to the interface between the black and white regions when they are closest to the LLM plane, which is the closest analog of the singularity in smooth solutions.

\subsubsection{Gray potential as average over black and white}

We now reach the core of the matter: are the grayscale orbits well approximated by ensemble averages over black and white? We will now explicitly define the ensembles, and then proceed to do the averaging.

For the whole sequence of $2N+1$ boundaries in a multiring background of Fig.~\ref{figgrayrings}(A) the total black and white potential reads\footnote{We adopt cylindrical coordinates as the expressions are simpler.}
\be
V_\mathrm{BW}(\xi)=\frac{1}{\xi^2}\sum_{j=1}^{2N+1}(-1)^{j+1}\left[J_-^2\Theta(\rho-R_j)+J_+^2\Theta(R_j-\rho)\right],\label{veffbwtot}
\ee
while the grayscale system has the contribution (\ref{veffbw}) for $0<\rho<R_1$ and $R_2<\rho$ and (\ref{veffgray}) for $R_1<\rho<R_2$:
\bea
V_\mathrm{gray}(\xi)&=&\frac{1}{\xi^2}\bigg[J_-^2\left(\Theta\left(\rho-R_1\right)+\Theta\left(\rho-R_3\right)\right)+J_+^2\left(\Theta\left(R_1-\rho\right)+\Theta\left(R_3-\rho\right)\right)-\nonumber\\
&&\left(\frac{E}{2}\right)^2\left(1-g^2\right)+\frac{J_-^2+J_+^2}{2}+\frac{g}{2}\left(J_-^2-J_+^2\right)\mathrm{sgn}\left(\rho-R_2\right)\bigg]\label{veffgraytot}
\eea
We now want to compare (\ref{veffgraytot}) to the ensemble-averaged value of (\ref{veffbwtot}). 

In principle, one should quantize the LLM system and average over the fluctuations of the ground state, similar to the procedure for the five-brane stars in \cite{Martinec:2023gte,Martinec:2023xvf}. We, however, find this too complicated, and also the global shape of the geodesics should be insensitive to the details of the in-plane black and white pattern, which only influence extremely low-energy behavior. We thus adopt a purely ad hoc Gaussian distribution of $R_2,\ldots R_{2N-1}$ (satisfying the constraint (\ref{grayconstraint})):
\be
P(R_2,\ldots R_{2N-2})=\mathcal{N}e^{-\sum_{j=2}^{2N-1}\frac{(R_j-R_{j;0})^2}{2\sigma^2}}\delta\left(\sum_{j=1}^{2N+1}R_j^2-\frac{\mathcal{A}_0}{\pi}\right)\label{rgauss}
\ee
where the mean values $R_{i;0}$ are taken to be uniformly spaced:
\bea
R_{j;0}&=&R_1+(j-1)\frac{R_{2N}-R_1}{N-2}\nonumber\\
\mathcal{A}_0&=&\pi\sum_{j=1}^{N}\left(R_{2j}^2-R_{2j-1}^2\right)
\eea
Now representing the Dirac delta through its Fourier transform the ensemble-averaged potential becomes
\bea
\langle V_\mathrm{eff}(\xi)\rangle&=&\frac{1}{Z}\frac{1}{\xi^2}\prod_{j=2}^{2N-1}\int dR_j\int d\lambda V_\mathrm{BW;eff}(\xi)\times\nonumber\\
&&\exp\left[-\sum_{j=2}^{2N-1}R_j^2\left(\frac{1}{2\sigma^2}+(-1)^j\imath\lambda\right)+\sum_{j=2}^{2N-1}\frac{R_{j;0}}{\sigma}R_j-\imath\lambda\Sigma_-\right],\label{veffaverage}
\eea
where we have dropped the constant factors as they cancel out with the same factors in the partition function $Z$, and we have denoted $\Sigma_-\equiv\sum_{j=1}^{2N+1}(-1)^{j+1}R_{j;0}^2$. We can write the above in matrix form as
\be
\langle V_\mathrm{eff}(\xi)\rangle=\frac{1}{Z}\frac{1}{\xi^2}\prod_{j=2}^{2N-1}\int dR_j\int d\lambda V_\mathrm{BW;eff}(\xi)\exp\left[-\mathbf{R}\cdot\hat{\mathbf{M}}\cdot\mathbf{R}-\mathbf{K}\cdot\mathbf{R}-\imath\lambda\Sigma_-\right],\label{veffaveragemat}
\ee
where the vector $\mathbf{R}\equiv(R_2,\ldots R_{2N-1})$ contains the random ring radii and the $(2N-2)\times (2N-2)$ matrix $\hat{\mathbf{M}}$ and the $(2N-2)$-vector $\mathbf{K}$ read
\be
\hat{\mathbf{M}}=\mathrm{diag}\left(\frac{1}{2\sigma^2}+(-1)^j\imath\lambda\right),~~\mathbf{K}=\left(\frac{R_{j;0}}{\sigma}\right),\quad j=2,\ldots 2N-1.\label{veffmk}
\ee
Inserting the expression (\ref{veffbw}) into (\ref{veffaveragemat}) we see that the step functions just modify the limits of the $R_j$-integrals as
\be
\int dR_j\int_0^\infty d'\mathbf{R}\mapsto\sum_{j=2}^{2N-1}\left(J_-^2\int_0^\rho dR_j+J_+^2\int_\rho^\infty dR_j\right)\int_0^\infty d'\mathbf{R},
\ee
where $\int d'\mathbf{R}\equiv\prod_{k\neq j}\int dR_k$ and in general $\mathbf{R}'=(R)_{k,k\neq j}$. This yields
\bea
\langle V_\mathrm{eff}(\xi)\rangle&=&\frac{1}{Z}\sum_{j=2}^{2N-1}\int d\lambda\bigg[\frac{J_-^2}{\xi^2}\int_0^\rho dR_j\int_0^\infty d'\mathbf{R}e^{-R_j^2\left(\frac{1}{2\sigma^2}-\imath\lambda\right)-\frac{R_{j;0}R_j}{\sigma}}e^{-\mathbf{R}'\cdot\hat{\mathbf{M}}'\cdot\mathbf{R}'-\mathbf{K}'\cdot\mathbf{R}'}-\nonumber\\
&&-\frac{J_+^2}{\xi^2}\int_\rho^\infty dR_j\int_0^\infty d'\mathbf{R}e^{-R_j^2\left(\frac{1}{2\sigma^2}-\imath\lambda\right)-\frac{R_{j;0}R_j}{\sigma}}e^{-\mathbf{R}'\cdot\hat{\mathbf{M}}'\cdot\mathbf{R}'-\mathbf{K}'\cdot\mathbf{R}'}\bigg]e^{-\imath\lambda\Sigma_-},\label{veffaverageint}
\eea
where, analogously to the definition of $\mathbf{R}'$, we have $\mathbf{K}'=(K_k)_{k\neq j}$ and $\mathbf{M}'=(M_{kl})_{k,l\neq j}$. Performing the integrals yields
\bea
&&\langle V_\mathrm{eff}(\xi)\rangle=\frac{1}{Z}\frac{\sqrt{\pi}}{2\xi^2}\int d\lambda\sum_{j=2}^{2N-1}\frac{(-1)^{j+1}}{\sqrt{M_{(-1)^j}}}e^{-\frac{R_{j;0}^2}{4\sigma^2M_{(-1)^j}}}\times\nonumber\\ &&\left[\left(J_-^2-J_+^2\right)\mathrm{Erf}\left(\frac{R_{j;0}+2\sigma M_{(-1)^j}\rho}{2\sigma\sqrt{M_{(-1)^j}}}\right)+J_+^2-J_-^2~\mathrm{Erf}\left(\frac{R_{j;0}}{2\sigma\sqrt{M_{(-1)^j}}}\right)\right]e^{-\imath\lambda\Sigma_-}Z_{2N-3;j}.~~~~~~~~\label{veffaverageint2}
\eea
Here, $\mathrm{Erf}$ is the error function and $Z_{2N-3;j}$ is the partition function for $2N-3$ disks with radii $R_k$ with $k\neq 1,j,2N,2N+1$; we can also denote $Z_{2N-2}\equiv Z$. The partition function is written out as
\bea
&&Z=Z_{2N-2}=\frac{\pi^{N-1}}{(M_+M_-)^{\frac{N-1}{2}}}\prod_{j=2}^{2N-1}e^{\frac{R_{j;0}^2}{4\sigma^2M_{(-1)^j}}}\left[1-\mathrm{Erf}\left(\frac{R_{j;0}}{2\sigma\sqrt{M_{(-1)^j}}}\right)\right]\label{zfull}\\
&&Z_{2N-3;j}=\frac{\pi^{N-1/2}}{(M_+M_-)^{\frac{N-2}{2}}M_{(-1)^k}^{1/2}}\prod_{k\neq j}e^{\frac{R_{k;0}^2}{4\sigma^2M_{(-1)^k}}}\left[1-\mathrm{Erf}\left(\frac{R_{k;0}}{2\sigma\sqrt{M_\pm}}\right)\right].\label{z2n3}
\eea
Noticing that
\be
Z_{2N-3;j}=\prod_{k\neq j}Z_{1;(k)},~~Z_{1;(k)}=\sqrt{\frac{\pi}{M_{(-1)^k}}}e^{\frac{R_{k;0}^2}{4\sigma^2M_\pm}}\left[1-\mathrm{Erf}\left(\frac{R_{k;0}}{2\sigma\sqrt{M_\pm}}\right)\right]\label{z1}
\ee
we get
\bea
\langle V_\mathrm{eff}(\xi)\rangle&=&\frac{1}{2\xi^2}\int d\lambda e^{-\imath\lambda\Sigma_-}\sum_{j=2}^{2N-1}\frac{A_j}{Z_{1;(j)}}\label{vefffin}\\
\frac{A_j}{Z_{1;(j)}}&=&\frac{\left(J_-^2+J_+^2\right)\mathrm{Erf}\left(x_j+\rho\sqrt{M_{(-1)^j}}\right)+\left(J_+^2-J_-^2\right)\mathrm{Erf}(x_j)}{1-\mathrm{Erf}(x_j)}\label{veffaz}\\
x_j&\equiv&\frac{R_{j;0}}{2\sigma\sqrt{M_{(-1)^j}}}.\label{veffaveragefin}
\eea
Now the remaining steps are the $\lambda$-integral and the coarse-graining. We do the integral in the saddle-point approximation, solving for the zeros off the integrand in (\ref{vefffin}), which can only be done numerically (recall that $x_j$ depends on $\lambda$ through $M_{(-1)^j}$ so the $\lambda$-derivative of the integrand is quite involved).

There is an infinite series of zeros; however, since $\lambda$ is really the Fourier conjugate variable of $R$ and $R$ is confined between $R_1$ and $R_{2N}$ it follows that $\lambda$ is periodic with period $2\pi/(R_{2N}-R_1)$; as usual, the saddle points outside the first Brillouin zone merely introduce a phase shift that cancels out. Calculating the residue and evaluating the integral we find a closed-form expression but very complicated and nonintuitive. We give it in Appendix \ref{secappvmean}.

\begin{figure}[ht]
    \centering
    (A)\includegraphics[width=.45\linewidth]{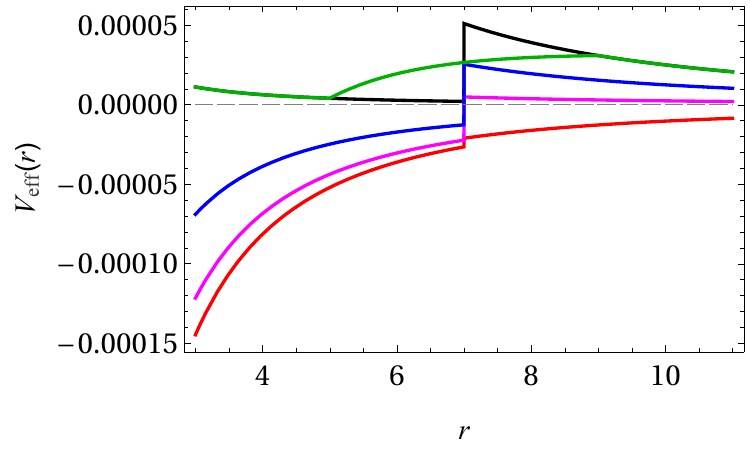}
    (B)\includegraphics[width=.45\linewidth]{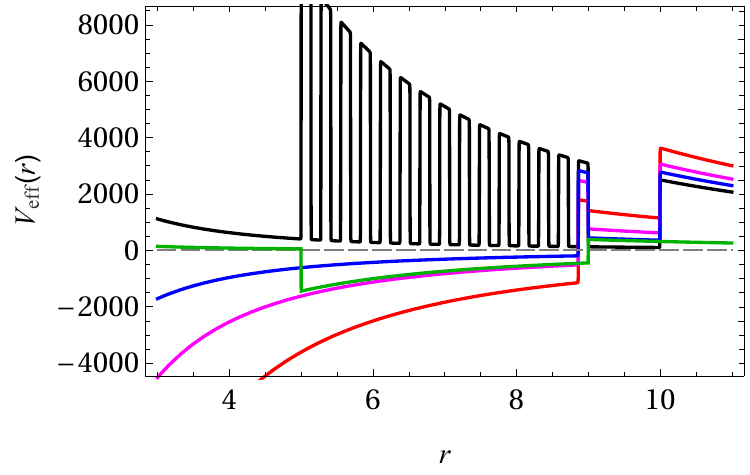}
    \caption{(A) Near-plane effective potential $V_\mathrm{eff}(r;\theta\mapsto 0)$ for a single black disk (black), single gray disk with decreasing flux $g=7/9,5/9,1/9$ (blue, magenta, red) and for the average over the size of the black disk (green). The integrals of motion are $(E,J_-,J_+,P_\phi)=(0.1,0.05,0.01,0.1)$. For a single disk the averaging does not mimic the grayscale geometry, i.e. does not create a potential well (though it does create metastable bound states). (B) Same as in (A) for disk+rings configuration with $2N+1=31$ rings. The averaged (green) curve now provides a good approximation of the grayscale potential with appropriate nuance $g=5/9$ (magenta). In particular, bound states and falling into the center are reproduced.}
    \label{figeffpots}
\end{figure}

The plots of the black and white, grayscale and ensemble-averaged potential as a function of $r,\theta$\footnote{Even though the expressions for a single disk are much simpler in polar coordinates, the integration is easier to do in spherical coordinates. This fact and the fact that cylindrical coordinates are not convenient for the 3-disk case have prompted us to use spherical coordinates.} in Fig.~\ref{figeffpots} reveal that the average introduces the potential well of the naked grayscale singularity when there are many rings to coarse-grain over (panel B). If we only average the position of a single disk (panel A), the average does not have true bound (negative energy) states.

Still, even in the single-disk case there is one important effect: the potential barrier, i.e. the strength of the repelling potential, is reduced (even though it remains positive, i.e. repelling). This effect will push the orbits toward grayscale behavior, \emph{on timescales which are not too long}, in fact shorter than the trapping timescale. In other words, precisely because of the repelling nature of the potential exactly at $\xi=0$, where the black/white pattern is strict, most geodesics cannot come very close to the LLM plane and for that reason do not differentiate much between grayscale and averaged black and white as far as motion in the $\xi$ direction is concerned. When many disks are present, the total effect is larger and we end up with a very good approximation of the grayscale potential, at just the right nuance for this numerical example $g=5/9$.

Finally, we note that conceptually similar result is obtained if we assume a uniform distribution of ring radii, as shown in Appendix \ref{secappuni}. This is less justified as an approximation to actual fluctuations of the background (Gaussian distribution being favored by the usual central limit theorem arguments) but the final result confirms the robustness of the averaging picture.

Another way to think about the universality of these behaviors is that the function $z$ which appears in all these geometries satisfies an elliptic PDE and is similar to a solution of the Laplace equation with fixed (Dirichlet) boundary conditions. The Laplace equation smoothes configurations rapidly away from the boundary, so the precise details of the structure that is exactly at the boundary becomes unimportant when we are some distance away. Also, notice that in all the black and white solutions the function $h$ grows as we try to reach a black and white interface. That is because in this region $z$ is farther away from $\pm 1/2$ as we approach the LLM plane.
This provides a reason for the wells to be forming, even though it is not obvious. 

\subsection{Averaged trajectories}

One could argue that an orbit \emph{in averaged effective potential} is not equivalent to an \emph{averaged orbit} itself, i.e. that averaging and solving the equations of motion do not commute. This is certainly true in the rigorous sense, however numerical integrations show that orbits in averaged potentials make a very good approximation of the orbit averages \emph{up to some timescale}.

\begin{figure}[ht]
    \centering
    (A)\includegraphics[width=.40\linewidth]{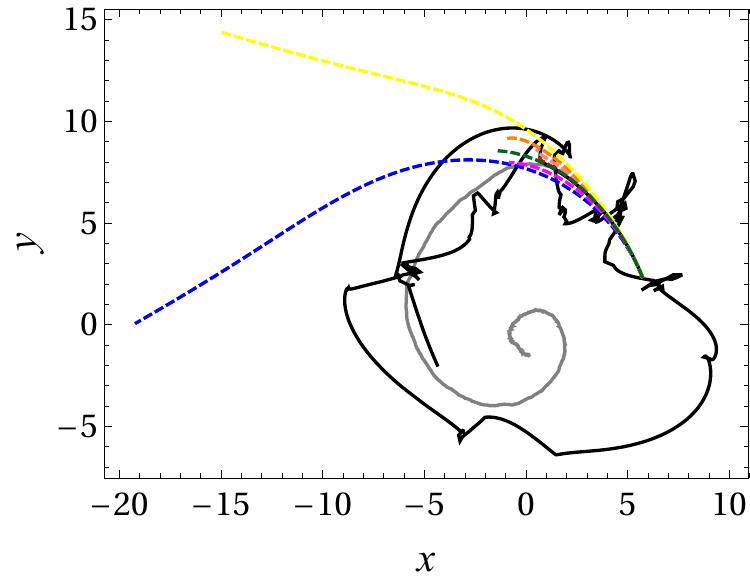}
    (B)\includegraphics[width=.49\linewidth]{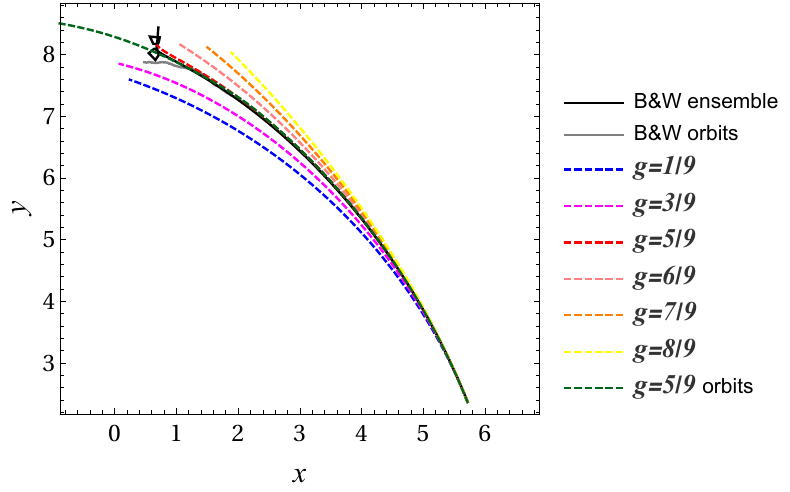}
    \caption{Black and white orbits averaged over the ensemble (black) or over the initial conditions (gray), compared to the grayscale orbits, including the one with the correct averaged flux (green, $g=5/9$), which indeed shows near-perfect self-averaging for short times.}
    \label{figorbitsaverage}
\end{figure}

In Fig.~\ref{figorbitsaverage} we compare the orbit in the averaged potential to the orbits in different grayscale backgrounds. For reference, we also include the phase space average, i.e. the average from different initial conditions in a fixed black and white background. The grayscale orbit approximates the ensemble average very well, precisely for the flux that equals the total flux in the black and white ensemble, which further confirms that the averaging works as expected. But this is true for early enough times, i.e. roughly until some crossover time $t_c$. For $t>t_c$ the agreement is nonexistent as the grayscale orbits end up in the singularity or at infinity -- in other words, either $t_\mathrm{esc}\sim t_c$ or $t_\mathrm{fall}\sim t_c$, whereas for the black and white backgrounds $t_\mathrm{esc}$ is much longer as we have seen earlier, and there is no singularity ($t_\mathrm{fall}=0$). This is in agreement with the general intuition that self-averaging works at short times. Finally, the fact the phase-space averaging is nearly equivalent to ensemble averaging of backgrounds shows that both kinds of averaging crucially depend on dynamical chaos: phase space averaging is always a signature of developed chaos (which again only works on timescales short enough that most orbits do not explore the remnants of KAM tori).

\subsection{Grayscale and averaging in 3-disk geometries}

It is instructive to apply the same tests on averaging and grayscale geometries to the 3-disk solution. We envision the following setup: one of the disks is a configuration of $2N+1$ rings of radii $0<R_1<\ldots R_{2N}<R_{2N+1}$ with fixed outer radius $R_{2N+1}=\mathrm{const.}$ and fixed total flux $\mathcal{A}$. The remaining two disks remain fully black, with radii $R_{2N+1}$, and their centers are located at the same points as in Fig.~\ref{figbackgnd3disks}. Now we go through the same numerical tests: we integrate geodesics on ensembles of fixed disks+rings backgrounds, then replace the $2N+1$-ring disk by a gray disk with the same flux and position (but necessarily different radius), and compare the distribution functions and orbits.

Averaged orbits (see Appendix \ref{secapp3disks}) show the same general picture as before, but now the self-averaging timescale is much longer. The reason is easy to trace to the absence of the extra integral of motion and thus lesser sensitivity of the system to quasiregular structures. 

It is tempting to conclude that the picture of a black hole as an average over microstates applies better to the 3-disk solution than to the disk+rings solution. However, the latter is much easier to interpret and much closer to an $AdS$ black hole metric, whereas the former is a rather exotic, anisotropic configuration.\footnote{Both have the same topology and the same boundary $\mathbb{R}\times\mathbb{S}^3$, as both have a final area of black regions and thus a zero total fraction of black regions in the LLM plane; in the classification of \cite{Mosaffa:2006qk}, this is the $\langle z\rangle=1/2$ case, as opposed to $\langle z\rangle<1/2$ geometries which have a lightlike boundary. But in our case it is not only the topology that is important, so the fact that the 3-disk and disk+ring solutions belong to the same class might not matter much.} This either means that the averaging we have found in the paper is nevertheless not closely related to black holes, or that quantum corrections would further reduce the measure of stability islands and sticky orbits and increase the averaging time also for the disk+ring configuration.

\section{Discussion and conclusions}\label{secconc}

We have embarked on this work hoping to achieve a coherent picture of how ensemble averaging works in higher-dimensional ($D>3$) gravity, and to understand if superpositions or coarse-graining of supersymmetric top-down solutions can approach the phenomenology of black holes.

Our main tool and paradigm is the study of null geodesics in these backgrounds and their chaotic dynamics: we know that black holes are dual to maximally chaotic field theories, and that they nearly always have integrable geodesic dynamics -- both facts stemming from the near-horizon and inner-region $SL(2,\mathbb{R})$ symmetries. Our analysis shows that black holes are but one extreme in the hierarchy: black holes $\rightarrow$ ``incipient black holes'' (naked singularities with no horizon but good in the Gubser sense) $\rightarrow$ smooth microstate geometries. Black holes mostly have trivial geodesic dynamics, grayscale solutions are nonintegrable but still show only weak chaos as most orbits are swallowed by the singularity before they develop chaos, and in smooth black and white geometries there is developed geodesic chaos and complex phase space with fractal structure. Therefore, presumably, the less chaotic bulk, so long as it has a singularity or a horizon,  the more chaotic the dual field theory.

Averaging over black and white backgrounds produces effective potentials with potential wells, which mimic the singularity in grayscale. This is somewhat counterintuitive -- the potential itself is obviously a highly non-self-averaging quantity as we average over potentials with no obvious deep wells and get a deep well as a result. It turned out that this idea of no deep wells is closer to failure exactly at the interfaces of black and white regions. When we have a grayscale geometry, these interfaces are populating the whole of the LLM plane and dominate the details of the geometry.
But that precisely makes the orbits themselves very much self-averaging, at least at not too long timescales. This is an explicit example where averaging over backgrounds in higher-dimensional gravity brings us closer to black hole dynamics.
We also found out that the time-time component of the metric vanishes in these singular solutions, so the averaged singularity is almost a horizon and the dynamics of infalling geodesics freezes near the singularity as viewed by an external observer.

Much remains to be done. First, the averaging over backgrounds that we perform is just phenomenological -- in fact one should integrate over the actual quantum fluctuations of LLM geometries. Second, it might make sense to look at non-asymptotically-$AdS$ LLM patterns, as the $AdS$ boundary acts as a potential box itself and influences the long time dynamics of the system. Third, one should understand the role of unstable periodic orbits, their behavior under averaging and if LLM solutions have an analog notion to a photon ring. Finally, the fractal structure of the bulk can tell us something about the behavior of the correlation functions in the eikonal approximation. We will address the last two points in separate works.

% Nonintegrability as expected; Chaos directly leads to the trapping; Much less developed chaos for grayscale, because of two attractors; Orbits i.e. the propagators are self-averaging

\acknowledgments

We are grateful to Andrei Parnachev, Jorge Russo, Yoav Zigdon and Anayeli Ramirez for stimulating discussions. D.B. work supported in part by
the Department of Energy under grant DE-SC 0011702. DB is also supported in part by the Delta ITP consortium,
a program of the Netherlands Organisation for Scientific Research (NWO) funded by the
Dutch Ministry of Education, Culture and Science (OCW) for the early portions of this work.
Work at the Institute of Physics is funded by the Ministry of Education, Science and Technological Development and by the Science Fund of the Republic of Serbia. The work on Sections 1 and 2 was supported by Russian Science Foundation Grant No. 24-72-10061 [https://rscf.ru/project/24-72-10061/] and performed at Steklov Mathematical Institute of Russian Academy of Sciences (Mihailo \v{C}ubrovi\'c).

%M.~\v{C}. would like to acknowledge the Mainz Institute for Theoretical Physics (MITP) of the Cluster of Excellence PRISMA+ (Project ID 39083149) for hospitality and partial support during the completion of this work.

\appendix 

\section{\label{secapplambda}Numerical calculation of the maximal Lyapunov exponent}

We will here present a method for numerical computation of the maximal Lyapunov exponent. More detailed elaboration can be found in \cite{Skokos:2008uc}. For a given solution to the Hamiltonian system $\vec{\mathcal X}(t)$ we can define a deviation vector $\delta \vec{\mathcal X}(t)$ that is a solution to the variational equations. We extract the maximal Lyapunov exponent by taking the following limit
\begin{equation}
    \lambda_1 = \lim_{t \rightarrow \infty} \frac{1}{t} \log \frac{|\delta \vec{\mathcal X}(t)|}{|\delta \vec{\mathcal X}(0)|}.
\end{equation}
This procedure may be difficult to implement due to limited memory resources, since $|\delta \vec{\mathcal X}(t)|$ can take very large values that are difficult to store on the computer. In such a situation one implements the following trick
\begin{equation} \label{LEtrick}
    \log \frac{|\delta \vec{\mathcal X}(N \Delta t)|}{|\delta \vec{\mathcal X}(0)|} = \log \left( \frac{|\delta \vec{\mathcal X}(N \Delta t)|}{|\delta \vec{\mathcal X}((N-1) \Delta t))|} \cdot \frac{|\delta \vec{\mathcal X}((N-1) \Delta t))|}{|\delta \vec{\mathcal X}((N-2) \Delta t))|} \cdots \frac{|\delta \vec{\mathcal X}(\Delta t)|}{|\delta \vec{\mathcal X}(0)|} \right),
\end{equation}
where $t = N \Delta t$. Therefore, instead of solving the variational equations once on the interval $t \in [0,~t]$, one solves it $N$ times on a sequence of much smaller intervals $t \in [(k-1)~ \Delta t,~ k~ \Delta t], ~ k = 1,2, \cdots, N$. From Eq. \ref{LEtrick} it is then obvious that the total value of the maximal Lyapunov exponent is related to a sum of the following sequence
\begin{equation}
    \log \frac{|\delta \vec{\mathcal X}(t)|}{|\delta \vec{\mathcal X}(0)|} = \sum_{k=1}^{N} \log \frac{|\delta \vec{\mathcal X}(k \Delta t)|}{|\delta \vec{\mathcal X}((k-1)\Delta t))|}.
\end{equation}
A follow-up question is related to how fast such a sequence will converge towards the maximal Laypunov exponent value as we vary the number of steps $N$. Suppose that we calculate the maximal Lyapunov exponent as a sequence labeled by a number of steps $N_i$: $\lambda_1(N_i)$. In order to extract the limiting value of such a sequence in a computationally efficient way, methods to improve the convergence, such as the Richardson extrapolation \cite{Bender:1999box}, can be used.

\section{\label{secapp3disks}Escape time, Pesin relation and fractal structure for the 3-disk system}

\begin{figure}[ht]
    \centering
    (A)\includegraphics[width=.44\linewidth]{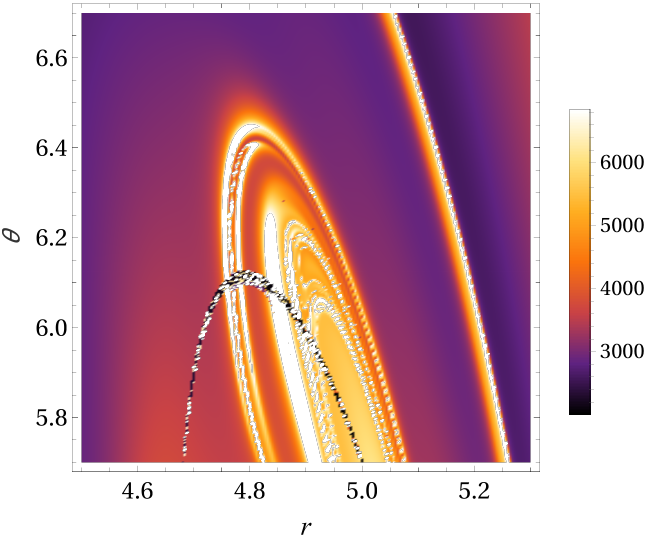}
    (B)\includegraphics[width=.47\linewidth]{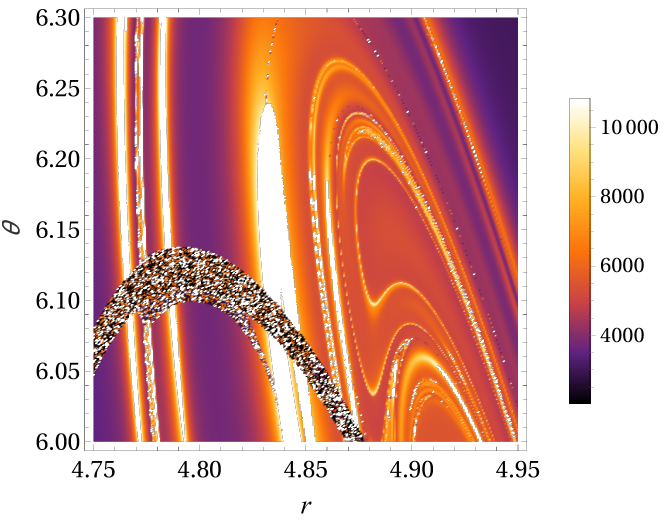}
    (C)\includegraphics[width=.47\linewidth]{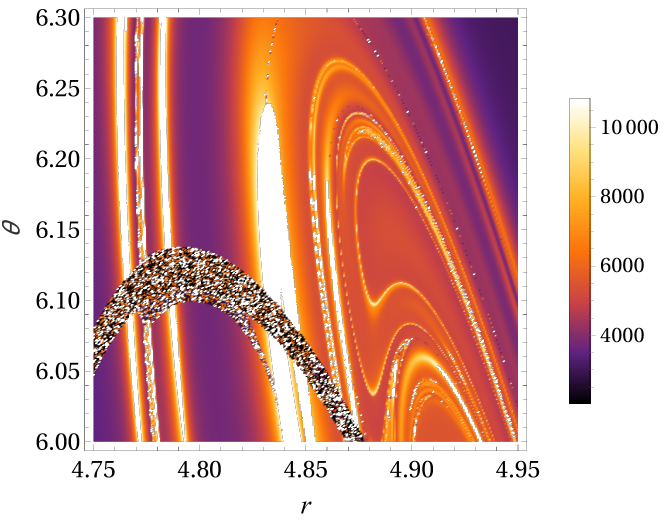}
    (D)\includegraphics[width=.43\linewidth]{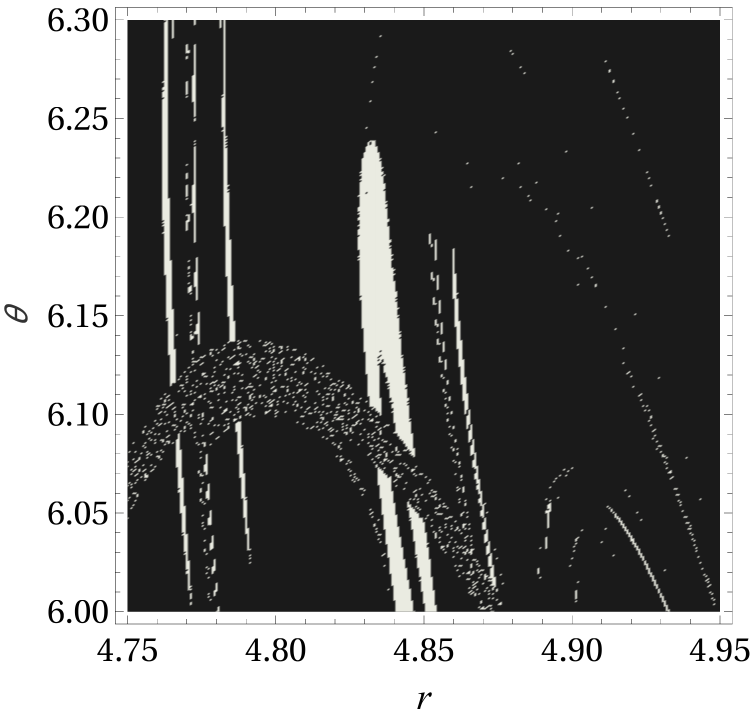}
    \caption{Escape time for a regular grid of initial conditions in $(\rho,\theta)$, with $P_\xi^{(0)}=-0.01,P_\rho^{(0)}=0.001$ and integrals of motion $(E,J_-,J_+)=(0.1,0.01,0.01)$. In (A-C) every initial condition along $\xi$ and $r$ is color-coded for escape time, at three levels of magnification. In (D) we indicate the sticky orbits ($t_\mathrm{esc}/t_0>50000$) at the same zoom level as in (B): these are the orbits with very long escape times in the vicinity of the remnants of KAM tori.}
    \label{figesctime3disk2d}
\end{figure}

\begin{figure}[ht]
    \centering
    \includegraphics[width=.85\linewidth]{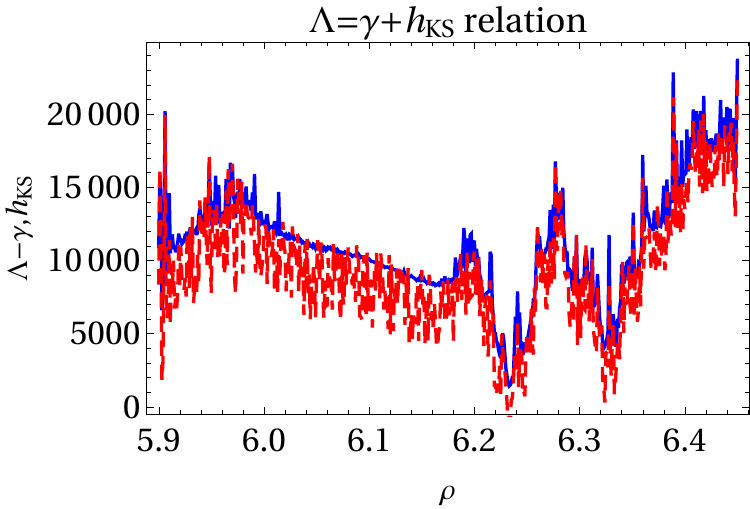}
    \caption{The numerical check of the Pesin relation (\ref{eqrelation}) between the KS entropy $h_\mathrm{KS}$, escape rate $\bar{\gamma}$ and the sum of positive Lyapunov exponents $\Lambda$, for the 3-disk configuration. We plot the KS entropy (red) versus the difference $\Lambda-\bar{\gamma}$ (blue) which is expected to equal the KS entropy. While the fluctuations are large, there is obviously a strong correlation. The calculations are done for the 3-disk system with the initial conditions $\xi^{(0)}=4.7,P_\xi^{(0)}=-0.01,\rho^{(0)}=6.15\pm 0.20,P_{\rho}^{(0)}=0.0010\pm 0.0001,\phi^{(0)}=\pi/8$ and integrals of motion $(E,J_-,J_+,P_\phi)=(0.1,0.01,0.01,0.002)$.}
    \label{figrelation3disk}
\end{figure}

\begin{figure}[ht]
    \centering
    (A)\includegraphics[width=.28\linewidth]{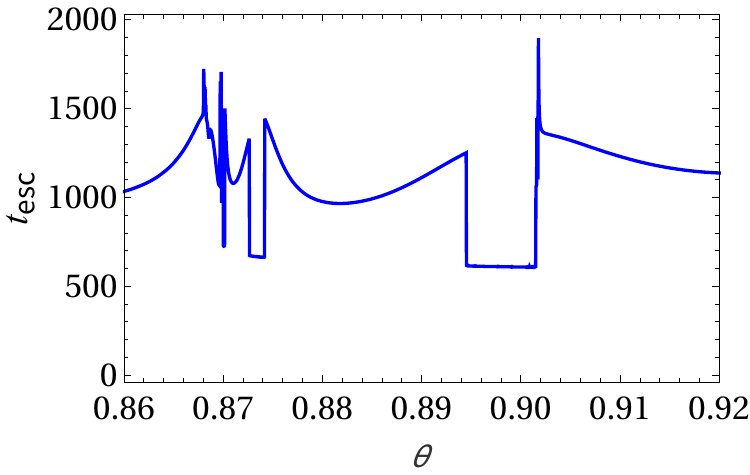}
    (B)\includegraphics[width=.28\linewidth]{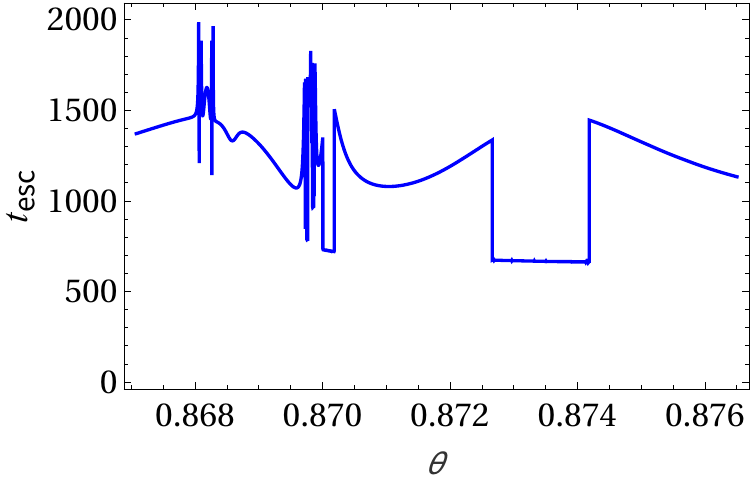}
    (C)\includegraphics[width=.28\linewidth]{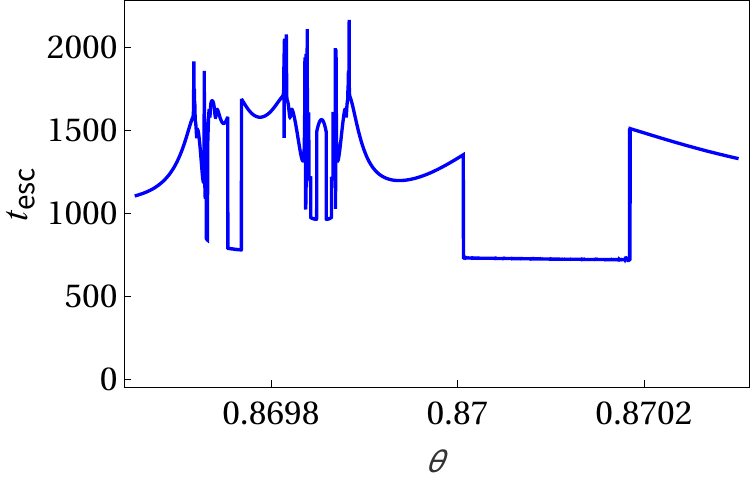}
    (D)\includegraphics[width=.28\linewidth]{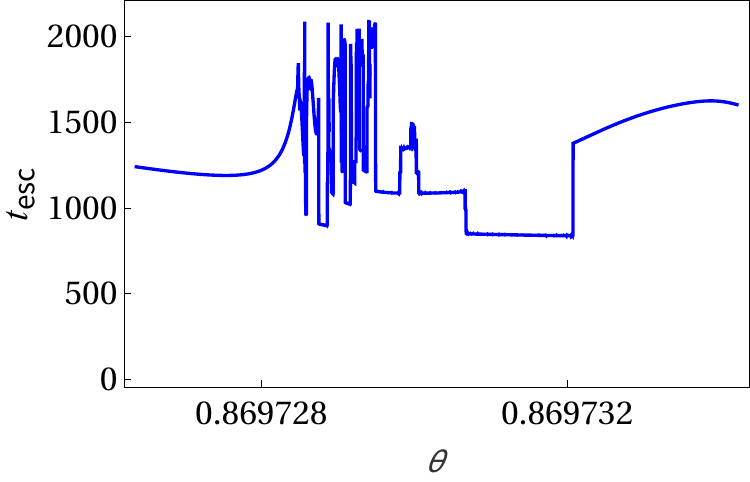}
    (E)\includegraphics[width=.28\linewidth]{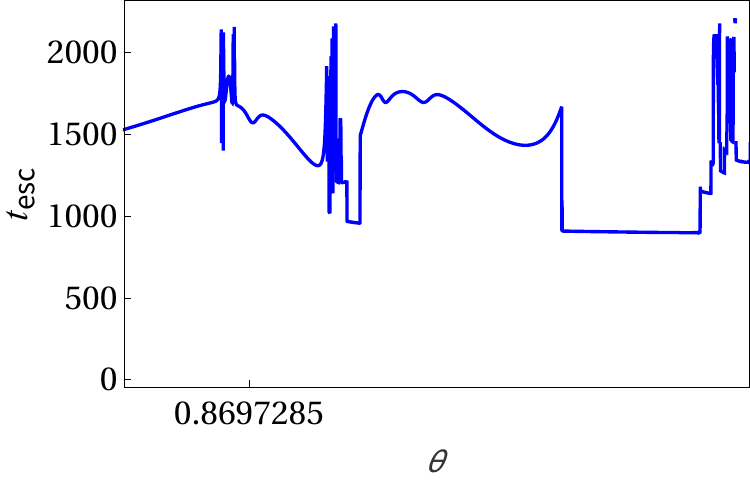}
    (F)\includegraphics[width=.28\linewidth]{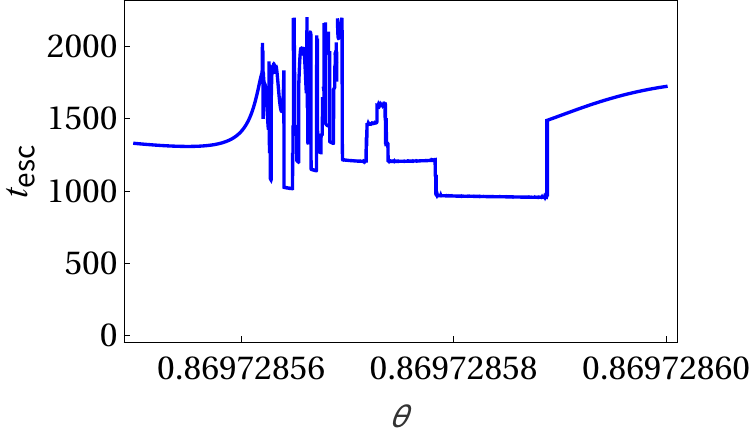}
    \caption{Escape time as a function of the angle $\theta$ at six zoom levels, showing approximately self-similar structures. The initial conditions and integrals of motion are the same as in Fig.~\ref{figesctime3disk2d} (what we show here are thus 1D sections of Fig.~\ref{figesctime3disk2d}).}
    \label{figesctime3disk1d}
\end{figure}

As we have commented in Section \ref{secbw}, the 3-disk system lacks the extra integral of motion of the disk+ring system, and thus has a higher relative measure of the chaotic component in the phase space. We have seen that this makes it more self-averaging then the simpler disk+ring system. But in terms of purely dynamic (as opposed to statistical) quantities, the phenomenology is similar. In this Appendix we show the figures analogous to Figs.~\ref{figesctime3disk2d}-\ref{figesctime3disk1d} for the disk+ring system and find an overall similar story.

We see this first in Fig.~\ref{figesctime3disk2d}, showing the escape rate at three levels of magnification (panels A-C), revealing the self-similar stability island structures. In panel (D) the sticky orbits (with very long escape times) are indicated, delineating the remnants of KAM tori.

Numerical check of the $\Lambda-h_\mathrm{KS}-\gamma$ relation for the 3-disk configuration is found in Fig.~\ref{figrelation3disk}.

The details of the fractal structure can be reproduced analytically to some extent, and they are related in an interesting way to the structure of two-point functions. However, this is a complex question in its own right and we address it in a separate work. Here, we just want to emphasize the typical structure of mixed chaotic-regular phase space found in LLM geodesics.

\section{\label{secappvmean}Explicit expression for the averaged potential with Gaussian fluctuations}

As we mention in the main text, closed-form expression for the integral exists but it is too long and nonilllustrative. We give it here for completeness.
\bea
\label{vefffinal}
&&V_\mathrm{eff}(r,\theta)=6 r \tanh ^{-1}\left(\frac{R_1}{r}\right) \left(J_-^2-J_+^2+4 E P_{\phi }\right)-6 r \tanh ^{-1}\left(\frac{R_{2
   N+1}}{r}\right) \left(J_-^2-J_+^2+4 E P_{\phi }\right)+\nonumber\\
   &&\frac{\left(R_1-R_{2 N+1}\right) \left(2 J_+^2
   \left(\left(\theta ^2-6\right) r^4+\left(3 R_1 R_{2 N+1} \theta ^2+2 \left(\theta ^2+3\right) R_1^2+2 \left(\theta
   ^2+3\right) R_{2 N+1}^2\right) r^2\right)\right)}{12r^2\theta ^2 \left(r^2-R_1^2\right) \left(r^2-R_{2 N+1}^2\right)}-\nonumber\\
   &&\frac{\left(R_1-R_{2 N+1}\right)\left(2 \left(\theta ^2+3\right) R_1^2 R_{2 N+1}^2-3 \theta ^2
   \left(r^2-R_1^2\right) \left(r^2-R_{2 N+1}^2\right) E^2\right)}{12r^2\theta ^2 \left(r^2-R_1^2\right) \left(r^2-R_{2 N+1}^2\right)}-\nonumber\\
   &&\frac{3\theta^2\left(4 P_{\phi } \left(r^2-R_1^2\right) \left(r^2-R_{2
   N+1}^2\right) E+2J_-^2 \left(r^4+R_1 R_{2 N+1} r^2\right)+4 P_{\phi }^2 \left(r^2-R_1^2\right) \left(r^2-R_{2N+1}^2\right)\right)}{12r^2\theta ^2 \left(r^2-R_1^2\right) \left(r^2-R_{2 N+1}^2\right)}\nonumber\\
   &&r<R_1\\
&&V_\mathrm{eff}(r,\theta)=\frac{12 R_1 \left(\Im\left(J_-\right)-i \Re\left(J_-\right)\right){}^2}{\theta ^2}-6 r \tanh
   ^{-1}\left(\frac{R_1}{r}\right) \left(J_-^2-J_+^2+4 E P_{\phi }\right)+\nonumber\\
   &&\frac{2 R_1 \left(-\left(2 r^2+R_1^2\right)
   J_-^2+6 E P_{\phi } \left(R_1^2-r^2\right)+J_+^2 \left(2 r^2+R_1^2\right)\right)}{R_1^2-r^2}+\frac{12 J_+^2 R_{2
   N+1}}{\theta ^2}+\nonumber\\
   &&\left(3 \left(R_1-R_{2 N+1}\right) E^2-12 P_{\phi } R_{2 N+1}
   E+6 r \tanh ^{-1}\left(\frac{R_{2 N+1}}{r}\right) \left(J_-^2-J_+^2+4 E P_{\phi }\right)\right)+\nonumber\\
   &&\frac{2\left(r^2 \left(R_1-3 R_{2 N+1}\right)-R_1 R_{2 N+1}^2\right) J_-^2+12 P_{\phi }^2 \left(R_1-R_{2 N+1}\right)
   \left(r^2-R_{2 N+1}^2\right)}{r^2-R_{2 N+1}^2}+\nonumber\\
   &&\frac{2J_+^2 \left(r^2\left(R_1+R_{2 N+1}\right)+R_{2 N+1}^2 \left(2 R_{2
   N+1}-R_1\right)\right)}{r^2-R_{2 N+1}^2}\nonumber\\
   &&R_1<r<R_{2N+1}\\
&&V_\mathrm{eff}(r,\theta)=-6 r \tanh ^{-1}\left(\frac{R_1}{r}\right) \left(J_-^2-J_+^2+4 E P_{\phi }\right)+6 r \tanh ^{-1}\left(\frac{R_{2
   N+1}}{r}\right) \left(J_-^2-J_+^2+4 E P_{\phi }\right)+\nonumber\\
   &&\frac{\left(R_1-R_{2 N+1}\right) \left(2 J_-^2
   \left(\left(\theta ^2-6\right) r^4+\left(3 R_1 R_{2 N+1} \theta ^2+2 \left(\theta ^2+3\right) R_1^2+2 \left(\theta
   ^2+3\right) R_{2 N+1}^2\right) r^2\right)\right)}{12r^2\theta ^2 \left(r^2-R_1^2\right) \left(r^2-R_{2 N+1}^2\right)}-\nonumber\\
   &&\frac{\left(R_1-R_{2 N+1}\right)\left(\left(2 \left(\theta ^2+3\right) R_1^2 R_{2 N+1}^2\right)-3 \theta ^2
   \left(\left(r^2-R_1^2\right) \left(r^2-R_{2 N+1}^2\right) E^2\right)\right)}{12r^2\theta ^2 \left(r^2-R_1^2\right) \left(r^2-R_{2 N+1}^2\right)}-\nonumber\\
   &&\frac{3\theta^2\left(4 P_{\phi } \left(r^2-R_1^2\right) \left(r^2-R_{2
   N+1}^2\right) E+2 J_+^2 \left(r^4+R_1 R_{2 N+1} r^2\right)+4 P_{\phi }^2 \left(r^2-R_1^2\right) \left(r^2-R_{2
   N+1}^2\right)\right)}{12r^2\theta ^2 \left(r^2-R_1^2\right) \left(r^2-R_{2 N+1}^2\right)}\nonumber\\
   &&r>R_{2N+1}.
\eea

\section{\label{secappuni}Averaging over black and white regions with a uniform distribution}

\begin{figure}[ht!]
    \centering
    \includegraphics[width=.65\linewidth]{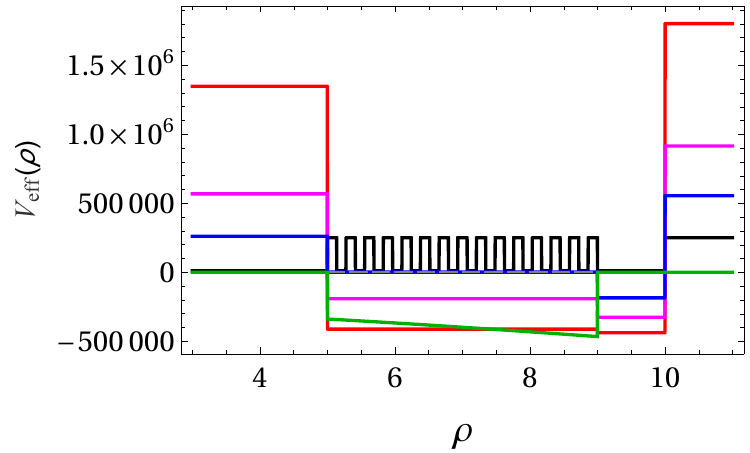}
    \caption{Near-plane effective potential $V_\mathrm{eff}(\rho;\xi\mapsto 0)$ for two black disks and one 31-ring configuration (black), two black disks and one gray with decreasing nuance $g=7/9,5/9,1/9$ (blue, magenta, red), and for the average over the rings (green). The integrals of motion are $(E,J_-,J_+,P_\phi)=(0.01,0.04,0.01,0.1)$. The averaged (green) curve qualitatively mimics the grayscale potentials but quantitative agreement is not as good as for the Gaussian case.}
    \label{figeffpotsuni}
\end{figure}

We will now repeat the averaging procedure for the effective potential from Eqs.~(\ref{rgauss}-\ref{veffaveragefin}) for a uniform distribution of the radii $R_{j;0}$, with $R_j\in\left[R_{j;0}-\Delta,R_{j;0}+\Delta\right]$. The probability distribution of the radii is then
\be
P(R_2,\ldots R_{2N-2})=\mathcal{N}\prod_{j=2}^{2N-1}\Theta\left(R_j-R_{j;0}+\Delta\right)\Theta\left(R_{j;0}+\Delta-R_j\right)\delta\left(\sum_{j=1}^{2N+1}R_j^2-\frac{\mathcal{A}_0}{\pi}\right).
\ee
In this case, it is most convenient to solve the constraint explicitly and eliminate $R_2$:
\be
R_2=\sqrt{\sum_{i=3}^{2N-2}{-1}^{i+1}R_i^2+R_1^2-R_{2N;0}^2+R_{2N+1}^2-\frac{\mathcal{A}}{\pi}},\label{veffr2}
\ee
where $\mathcal{A}$ is the (fixed) total area of gray rings, as before. Averaging the effective potential (\ref{veffbwtot}) in a manner analogous to Eqs.~(\ref{veffaverageint}-\ref{veffaverageint2}) now yields much simpler integrals (it is now more convenient to use polar coordinates):
\be
\langle V_\mathrm{eff}(\xi)\rangle=\frac{(2\Delta)^{3-2N}}{\mathcal{N}}\frac{J_-}{\xi^2}\sum_{i=2}^{2N-1}(-1)^i\prod_{j=3}^{2N-1}\int_{R_{j;-}}^{R_{j;+}}\left(\Theta(\rho-R_{i;0})+\Theta(R_{i;0}-\rho)\right)~+~(+~\leftrightarrow ~-),\label{veffaverageuni}
\ee
where $(+~\leftrightarrow ~-)$ denotes the part proportional to $J_+^2$, and $R_{i;\pm}=R_{i;0}\pm\Delta$. The integrals are now trivial, except for the term $i=2$ which contains the $R_j$-dependent function due to the solved constraint (\ref{veffr2}). This term evaluates to
\bea
\langle V_\mathrm{eff;2}(\xi)\rangle&\equiv&\frac{(2\Delta)^{3-2N}}{\mathcal{N}}\frac{J_-}{\xi^2}\prod_{j=4}^{2N-1}\int_{R_{j;-}}^{R_{j;+}}\int_{R_{3-}}^{\tilde{R}_{3+}}dR_3\nonumber\\
\tilde{R}_{3+}&\equiv&\mathrm{min}\left(R_{3+},\sqrt{\sum_{i=4}^{2N-2}{-1}^{i+1}R_i^2+\rho^2-R_1^2+R_{2N;0}^2-R_{2N+1}^2+\frac{\mathcal{A}}{\pi}}\right).
\eea
This integral evaluates in terms of trigonometric functions, resulting in very long expressions. However, if we assume that $\Delta\ll R_1$, which is logical as fluctuations should be smaller than the macroscopic scale of the background, we get a tractable expression:
\bea
\langle V_\mathrm{eff}(\xi)\rangle=\frac{1}{2\Delta\xi^2}\Bigg[(J_+^2-J_-^2)\sum_{i=3}^{2N-2}(-1)^{i}R_i+(J_+^2+J_-^2)\Delta-\Delta^2(J_+^2+J_-^2)\rho\Theta(\rho-R_1)\Theta(R_{2N}-\rho)-\nonumber\\
2N\Delta(J_+^2+J_-^2)\sqrt{\sum_{i=4}^{2N-2}{-1}^{i+1}R_i^2+\rho^2-R_1^2+R_{2N;0}^2-R_{2N+1}^2+\frac{\mathcal{A}}{\pi}}\Theta(\rho-R_1)\Theta(R_{2N}-\rho)\Bigg].\nonumber\\
\eea
The result is shown in Fig.~\ref{figeffpotsuni}. Now the averaged potential roughly corresponds to the grayscale ppotential at $g=1/9$, instead of the correct value $g=5/9$ for our setup. Therefore, even though we reproduce the necessary qualitative features, such as the potential well, the quantiative agreement is worse. Apparently, the microscopic dynamics of the fluctuations plays some role, and it is presumably closer to the Gaussian distribution as in most field theories.

\bibliography{LLM.bib}

\providecommand{\href}[2]{#2}\begingroup\raggedright\begin{thebibliography}{10}

\bibitem{Lin:2004nb}
H.~Lin, O.~Lunin, and J.~M. Maldacena, {\it {Bubbling AdS space and 1/2 BPS geometries}},  {\em JHEP} {\bf 10} (2004) 025, [\href{http://arxiv.org/abs/hep-th/0409174}{{\tt hep-th/0409174}}].

\bibitem{Berenstein:2004kk}
D.~Berenstein, {\it {A Toy model for the AdS / CFT correspondence}},  {\em JHEP} {\bf 07} (2004) 018, [\href{http://arxiv.org/abs/hep-th/0403110}{{\tt hep-th/0403110}}].

\bibitem{Bena:2004jw}
I.~Bena and N.~P. Warner, {\it {A Harmonic family of dielectric flow solutions with maximal supersymmetry}},  {\em JHEP} {\bf 12} (2004) 021, [\href{http://arxiv.org/abs/hep-th/0406145}{{\tt hep-th/0406145}}].

\bibitem{Mosaffa:2006qk}
A.~E. Mosaffa and M.~M. Sheikh-Jabbari, {\it {On classification of the bubbling geometries}},  {\em JHEP} {\bf 04} (2006) 045, [\href{http://arxiv.org/abs/hep-th/0602270}{{\tt hep-th/0602270}}].

\bibitem{Skenderis:2008qn}
K.~Skenderis and M.~Taylor, {\it {The fuzzball proposal for black holes}},  {\em Phys. Rept.} {\bf 467} (2008) 117--171, [\href{http://arxiv.org/abs/0804.0552}{{\tt arXiv:0804.0552}}].

\bibitem{Myers:2001aq}
R.~C. Myers and O.~Tafjord, {\it {Superstars and giant gravitons}},  {\em JHEP} {\bf 11} (2001) 009, [\href{http://arxiv.org/abs/hep-th/0109127}{{\tt hep-th/0109127}}].

\bibitem{Mandal:2005wv}
G.~Mandal, {\it {Fermions from half-BPS supergravity}},  {\em JHEP} {\bf 08} (2005) 052, [\href{http://arxiv.org/abs/hep-th/0502104}{{\tt hep-th/0502104}}].

\bibitem{Balasubramanian:2018yjq}
V.~Balasubramanian, D.~Berenstein, A.~Lewkowycz, A.~Miller, O.~Parrikar, and C.~Rabideau, {\it {Emergent classical spacetime from microstates of an incipient black hole}},  {\em JHEP} {\bf 01} (2019) 197, [\href{http://arxiv.org/abs/1810.13440}{{\tt arXiv:1810.13440}}].

\bibitem{Saad:2019lba}
P.~Saad, S.~H. Shenker, and D.~Stanford, {\it {JT gravity as a matrix integral}},  \href{http://arxiv.org/abs/1903.11115}{{\tt arXiv:1903.11115}}.

\bibitem{Blommaert:2020seb}
A.~Blommaert, {\it {Dissecting the ensemble in JT gravity}},  {\em JHEP} {\bf 09} (2022) 075, [\href{http://arxiv.org/abs/2006.13971}{{\tt arXiv:2006.13971}}].

\bibitem{Saad:2021rcu}
P.~Saad, S.~H. Shenker, D.~Stanford, and S.~Yao, {\it {Wormholes without averaging}},  {\em JHEP} {\bf 09} (2024) 133, [\href{http://arxiv.org/abs/2103.16754}{{\tt arXiv:2103.16754}}].

\bibitem{Blommaert:2021etf}
A.~Blommaert and M.~Usatyuk, {\it {Microstructure in matrix elements}},  {\em JHEP} {\bf 09} (2022) 070, [\href{http://arxiv.org/abs/2108.02210}{{\tt arXiv:2108.02210}}].

\bibitem{Blommaert:2021gha}
A.~Blommaert and J.~Kruthoff, {\it {Gravity without averaging}},  {\em SciPost Phys.} {\bf 12} (2022), no.~2 073, [\href{http://arxiv.org/abs/2107.02178}{{\tt arXiv:2107.02178}}].

\bibitem{Blommaert:2021fob}
A.~Blommaert, L.~V. Iliesiu, and J.~Kruthoff, {\it {Gravity factorized}},  {\em JHEP} {\bf 09} (2022) 080, [\href{http://arxiv.org/abs/2111.07863}{{\tt arXiv:2111.07863}}].

\bibitem{Frolov:2017kze}
V.~P. Frolov, P.~Krtous, and D.~Kubiznak, {\it {Black holes, hidden symmetries, and complete integrability}},  {\em Living Rev. Rel.} {\bf 20} (2017), no.~1 6, [\href{http://arxiv.org/abs/1705.05482}{{\tt arXiv:1705.05482}}].

\bibitem{Chervonyi:2013eja}
Y.~Chervonyi and O.~Lunin, {\it {(Non)-Integrability of Geodesics in D-brane Backgrounds}},  {\em JHEP} {\bf 02} (2014) 061, [\href{http://arxiv.org/abs/1311.1521}{{\tt arXiv:1311.1521}}].

\bibitem{Bena:2017upb}
I.~Bena, D.~Turton, R.~Walker, and N.~P. Warner, {\it {Integrability and Black-Hole Microstate Geometries}},  {\em JHEP} {\bf 11} (2017) 021, [\href{http://arxiv.org/abs/1709.01107}{{\tt arXiv:1709.01107}}].

\bibitem{Guica:2008mu}
M.~Guica, T.~Hartman, W.~Song, and A.~Strominger, {\it {The Kerr/CFT Correspondence}},  {\em Phys. Rev. D} {\bf 80} (2009) 124008, [\href{http://arxiv.org/abs/0809.4266}{{\tt arXiv:0809.4266}}].

\bibitem{Castro:2010fd}
A.~Castro, A.~Maloney, and A.~Strominger, {\it {Hidden Conformal Symmetry of the Kerr Black Hole}},  {\em Phys. Rev. D} {\bf 82} (2010) 024008, [\href{http://arxiv.org/abs/1004.0996}{{\tt arXiv:1004.0996}}].

\bibitem{Cvetic:2011hp}
M.~Cvetic and F.~Larsen, {\it {Conformal Symmetry for General Black Holes}},  {\em JHEP} {\bf 02} (2012) 122, [\href{http://arxiv.org/abs/1106.3341}{{\tt arXiv:1106.3341}}].

\bibitem{Cvetic:2011dn}
M.~Cvetic and F.~Larsen, {\it {Conformal Symmetry for Black Holes in Four Dimensions}},  {\em JHEP} {\bf 09} (2012) 076, [\href{http://arxiv.org/abs/1112.4846}{{\tt arXiv:1112.4846}}].

\bibitem{Chen:2024oqv}
Y.~Chen, H.~W. Lin, and S.~H. Shenker, {\it {BPS chaos}},  {\em SciPost Phys.} {\bf 18} (2025), no.~2 072, [\href{http://arxiv.org/abs/2407.19387}{{\tt arXiv:2407.19387}}].

\bibitem{Maldacena:2015waa}
J.~Maldacena, S.~H. Shenker, and D.~Stanford, {\it {A bound on chaos}},  {\em JHEP} {\bf 08} (2016) 106, [\href{http://arxiv.org/abs/1503.01409}{{\tt arXiv:1503.01409}}].

\bibitem{Djukic:2023dgk}
V.~Djuki\'c and M.~\v{C}ubrovi\'c, {\it {Correlation functions for open strings and chaos}},  {\em JHEP} {\bf 04} (2024) 025, [\href{http://arxiv.org/abs/2310.15697}{{\tt arXiv:2310.15697}}].

\bibitem{Berenstein:2023vtd}
D.~Berenstein, E.~Maderazo, R.~Mancilla, and A.~Ramirez, {\it {Chaotic LLM billiards}},  {\em JHEP} {\bf 08} (2024) 056, [\href{http://arxiv.org/abs/2305.19321}{{\tt arXiv:2305.19321}}].

\bibitem{Berenstein:2002jq}
D.~E. Berenstein, J.~M. Maldacena, and H.~S. Nastase, {\it {Strings in flat space and pp waves from N=4 superYang-Mills}},  {\em JHEP} {\bf 04} (2002) 013, [\href{http://arxiv.org/abs/hep-th/0202021}{{\tt hep-th/0202021}}].

\bibitem{Berenstein:2020jen}
D.~Berenstein and A.~Holguin, {\it {Open giant magnons on LLM geometries}},  {\em JHEP} {\bf 01} (2021) 080, [\href{http://arxiv.org/abs/2010.02236}{{\tt arXiv:2010.02236}}].

\bibitem{Eckmann:1985}
J.~P. Eckmann and D.~Ruelle, {\it Ergodic theory of chaos and strange attractors},  {\em Rev. Mod. Phys.} {\bf 57} (Jul, 1985) 617--656.

\bibitem{Berenstein:2016zgj}
D.~Berenstein and D.~Kawai, {\it {Smallest matrix black hole model in the classical limit}},  {\em Phys. Rev. D} {\bf 95} (2017), no.~10 106004, [\href{http://arxiv.org/abs/1608.08972}{{\tt arXiv:1608.08972}}].

\bibitem{Pesin:1977}
Y.~B. Pesin, {\it {Lyapunov characteristic exponents and smooth ergodic theory}},  {\em Usp. Mat. Nauk.} {\bf 32} (1977), no.~4 196.

\bibitem{Seoane:2013}
J.~M. Seoane and M.~A.~F. Sanjuán, {\it New developments in classical chaotic scattering},  {\em Reports on Progress in Physics} {\bf 76} (dec, 2012) 016001.

\bibitem{Caldarelli:2004mz}
M.~M. Caldarelli, D.~Klemm, and P.~J. Silva, {\it {Chronology protection in anti-de Sitter}},  {\em Class. Quant. Grav.} {\bf 22} (2005) 3461--3466, [\href{http://arxiv.org/abs/hep-th/0411203}{{\tt hep-th/0411203}}].

\bibitem{Gubser:2000nd}
S.~S. Gubser, {\it {Curvature singularities: The Good, the bad, and the naked}},  {\em Adv. Theor. Math. Phys.} {\bf 4} (2000) 679--745, [\href{http://arxiv.org/abs/hep-th/0002160}{{\tt hep-th/0002160}}].

\bibitem{Berenstein:2017abm}
D.~Berenstein and A.~Miller, {\it {Superposition induced topology changes in quantum gravity}},  {\em JHEP} {\bf 11} (2017) 121, [\href{http://arxiv.org/abs/1702.03011}{{\tt arXiv:1702.03011}}].

\bibitem{Martinec:2023gte}
E.~J. Martinec and Y.~Zigdon, {\it {BPS fivebrane stars. Part II. Fluctuations}},  {\em JHEP} {\bf 02} (2024) 034, [\href{http://arxiv.org/abs/2311.09157}{{\tt arXiv:2311.09157}}].

\bibitem{Martinec:2023xvf}
E.~J. Martinec and Y.~Zigdon, {\it {BPS fivebrane stars. Part I. Expectation values of observables}},  {\em JHEP} {\bf 02} (2024) 033, [\href{http://arxiv.org/abs/2311.09155}{{\tt arXiv:2311.09155}}].

\bibitem{Skokos:2008uc}
C.~Skokos, {\it {The Lyapunov Characteristic Exponents and their computation}},  {\em Lect. Notes Phys.} {\bf 790} (2010) 63--135, [\href{http://arxiv.org/abs/0811.0882}{{\tt arXiv:0811.0882}}].

\bibitem{Bender:1999box}
C.~M. Bender and S.~A. Orszag, {\em {Advanced Mathematical Methods for Scientists and Engineers I}}.
\newblock Springer, 1999.

\end{thebibliography}\endgroup

\end{document}